# Long term aging of Selenide glasses: Evidence of sub-$T_g$ endotherms and pre-$T_g$ exotherms

*Ping Chen\*, P. Boolchand*
Department of Electrical and Computer Engineering, University of Cincinnati, Cincinnati, OH 45221-0030

*D. G. Georgiev*
Department of Electrical Engineering and Computer Science, University of Toledo, Toledo, OH 43606

Long term aging, extending from months to several years, is studied on several families of chalcogenide glasses including the Ge-Se, As-Se, and Ge-As-Se systems. Special attention is given to the As-Se binary, a system that displays a rich variety of aging behavior intimately tied to sample synthesis conditions and the ambient environment in which samples are aged. Calorimetric (Modulated DSC) and Raman scattering experiments are undertaken. Our results show all samples display a *sub-$T_g$ endotherm* typically 10°C to 70°C below $T_g$ in glassy networks possessing a mean coordination number *r* in the 2.25 < *r* < 2.45 range. Two sets of $As_xSe_{100-x}$ samples aged for 8 years were compared, *set A* consisted of slow cooled samples aged in the dark, and *set B* consisted of melt quenched samples aged at laboratory environment. Samples of *set B* in the As concentration range, 35% < x < 60%, display a pre-$T_g$ *exotherm*, but the feature is not observed in samples of *set A*. The aging behavior of *set A* presumably represents *intrinsic* aging in these glasses, while that of set B is *extrinsic* due to presence of light. The reversibility window persists in both sets of samples but is less well defined in set B. These findings contrast with a recent study by Golovchak et al., which finds the onset of the reversibility window moved up to the stoichiometric composition ( x = 40%).  Here we show that the upshifted window is better understood as resulting due to demixing of $As_4Se_4$ and $As_4Se_3$ molecules from the backbone, i.e., Nanoscale phase separation (NSPS). We attribute *sub-$T_g$ endotherms* to compaction of the flexible part of networks upon long term aging, while the *pre-$T_g$ exotherm* to NSPS.  The  narrowing and sharpening of the reversibility window upon aging is interpreted  as the slow '*self-organizing*' stress relaxation of the phases just outside the Intermediate phase, which itself is stress-free and displays little aging.

## I.  INTRODUCTION

Structural glasses are intrinsically non-equilibrium solids. As the high liquid state entropy is dissipated upon undercooling and structural arrest manifests near T < $T_g$, the glass transition temperature, glasses continue to evolve with waiting time ($t_w$) extending up to months and years. We do not understand well the microscopic nature of entropy sinks that contribute to a reduction in entropy of a glass upon long term aging. Structural relaxations in glasses have been widely examined[1] and seem to follow a stretched exponential relaxation (SER) function, i.e., $(\exp -t/\tau)^\beta$ , where β is the stretching  exponent and τ , a characteristic time. Already in 1847, Kohlrausch[2] noted that the charge on a gold leaf electroscope in a Leyden jar follows a SER.  Different mechanisms have been advanced[1,3-6] for SER. The Scher-Lax-Phillips model of SER is based on the premise that either carriers and /or structure elements diffuse to traps and finds the stretched exponent, β = d*/(d*+2), where d* is an effective dimensionality.  Phillips[1] has emphasized the fact that two characteristic values of  β (3/5, 3/7) or effective dimensionalities d*(3,3/2) are found in a wide range of experiments in a wide array of materials systems [1,7].

The molecular substructure of network glasses has been key to understanding many features of 'fresh' glasses, i.e., non-aged. We are still far from understanding how aging occurs. Questions include: the nature of entropy sinks in real glasses that serve as traps in the SER model; the role of initial conditions and aging environment and the kind of long time structural changes involved. In this work we address some of these issues for an important class of glasses – the chalcogenides. Our approach is to examine long term aging of these systems at room temperature < $T_g$ as a function of waiting time ($t_w$) and use <u>m</u>odulated <u>D</u>ifferential <u>S</u>canning <u>C</u>alorimetry (m-DSC) along with Raman scattering as experimental probes. In the calorimetric experiments, we find evidence of sub-$T_g$ *endotherms* and in light exposed As-Se glasses pre-$T_g$ exotherms upon long term aging. The structural manifestations of these thermal events are deduced from Raman vibrational density of states.

Of particular interest is aging in intermediate phase (IP) glasses. Inside the IP narrow compositional windows the non-reversing enthalpy ($\Delta H_{nr}$) at $T_g$ becomes nearly vanishing compared to much larger values outside[8,9]. In addition, IP glasses are nearly stress free[10] at the molecular level, with a self-organized structure which makes them well suited for analyzing the dynamics of aging. In the IP of chalcogenides very little aging changes[10-13] are seen. Recently, Golovchak et al.[14] measured the non-reversing enthalpy at $T_g$ in $As_xSe_{100-x}$ samples aged for 22 years, and found the onset to begin at the stoichiometric composition, x = 40%. The apparent disagreement with our earlier results of an IP in the 29% < x < 37% range in 3 week aged glasses [15] prompted the present re-examination of that alloy system as well. The experimental results are presented in section II. A discussion of these results follows in section III and conclusions are given in section IV.

**II. EXPERIMENTAL**

### A. Sample synthesis, handling and characterization

Bulk chalcogenide glasses were synthesized using small lumps of 99.999% elemental Ge, Se, P, As and $As_2Se_3$ from Cerac Inc. as starting materials[11,15-17]. Elements in the desired ratio were sealed in evacuated ($10^{-7}$ Torr) quartz ampoules of 5mm id and reacted typically for 3 to 4 days at a suitably high temperature. Melt temperatures were then lowered to 50°C above the liquidus [18] kept there for a few hours, and then quenched in cold water. Sample homogeneity was established by comparing Raman spectra at dozen locations along the length of the quartz tube used for synthesis. Such Raman imaging experiments reveal that samples sealed at $10^{-7}$ Torr take much longer to react and homogenize[17] than those sealed at $10^{-5}$ Torr. If samples were found inhomogeneous, they were reacted for longer periods, and in most instances we could obtain homogeneous samples.

Modulated DSC experiments used a model 2920 instrument from TA Instruments Inc. About 20 mg quantity of a sample in a platelet form was hermetically sealed in Al pans. We used typically a 3°C/min scan rate and a 1°C/100sec modulation rate. We provide in Appendix 1 a brief overview of the method [19,20]. In Se-rich glasses, $T_g$s generally narrow upon aging, and even lower scan rates (0.1°C/min) were necessary. In m-DSC, $T_g$ is established[19,20] from the inflexion point of the rounded step observed in the reversing heat flow. The non-reversing enthalpy at $T_g$ was obtained by integrating the heat flow over the sub-$T_g$ and the $T_g$ endotherms, and subtracting off the exotherm in the cooling cycle to obtain the frequency corrected non-reversing enthalpy as discussed in Appendix 1[19,20].

An FT-Raman system (Thermo-Nicolet NEXUS 870) making use of 1064 nm radiation from a Nd:YAG laser was used to excite the scattering. A typical measurement used about 260mW of laser power brought to a loose focus of 400μm spot size on a sample, and the scattered light detected with either a Ge detector or InGaAs detector. A typical measurement used 200 scans for a 2 cm$^{-1}$ resolution.

For the case of binary As-Se glass system we will present work on *four sets* (**O, A, B, C**) of samples defined as follows. D.Georgiev et al.[15] synthesized melt-quenched $As_xSe_{100-x}$ glass samples over a wide range, 0 < x < 60%, in the year 2000. These samples were aged at room temperature for 3 weeks in quartz tubes used to synthesize them prior to their investigation in mDSC experiments. This set of samples were opened on a lab bench and transferred into hermetically sealed Al pans and calorimetric scans initiated followed by a cooling cycle back to RT , both usually at a scan rate of 3°C/min unless indicated otherwise. Results on these samples, henceforth labeled as **set O** (original) were reported upon in the year 2000[15] and the thermal and storage history illustrated in Fig. 1. These samples, slow cooled from $T_g$ and aged for 8 years in the original hermetically sealed Al pans, were re-investigated in 2008, and represent **set A** (Fig. 1). These samples neither saw light nor a laboratory humid environment for eight years of aging. The majority of samples synthesized by Georgiev et al.[15] in the year 2000 were stored in translucent plastic vials with slip fit caps at laboratory ambient environment, and saw natural light and laboratory humid environment over the 8 year aging period, these samples belong to **set B** (Fig. 1). It is useful to mention that samples of **set A** were slow cooled from $T_g$ while those of **set B** were melt-quenched prior to the 8-year aging process. Samples were then re- cycled through $T_g$, i.e., rejuvenated, and are denoted as belonging to **set C** (Fig. 1).

Aging results on several other glass systems were investigated as a function of $t_w$ at laboratory temperature for periods ranging from months to years. Details on synthesis of these samples can be found elsewhere[11,16]. All glass samples in these other studies were aged in hermetically sealed Al pans at room temperature.

### B. Experimental Results

#### 1. $Ge_xSe_{100-x}$

In Fig. 2 (a) and (b) we summarize aging results on a binary glass, $Ge_{16}Se_{84}$ examined after 4 weeks and 1 year of waiting time. Although the glass transition remains steady at 143°C, note that the sub-$T_g$ endotherm centroid moves up in T from about 100°C to about 120°C to coalesce with the $T_g$ endotherm.

#### 2. $Ge_xAs_xSe_{100-2x}$

Fig. 2(c) and (d) reproduce calorimetric data on the titled ternary at a composition x = 13% recorded after a waiting time of 6 months and then after 3 years. A perusal of these data shows the sub-$T_g$ endotherm centroid shifts from about 100°C to 110°C. The sub-$T_g$ endotherm in the ternary at x = 13% (Fig. 2(c) and (d)) is much smaller in size than in the $Ge_xSe_{100-x}$ binary at x = 16% (Fig. 2 (a) and (b)). The former corresponds to a mean coordination number r = 2 + 3 x = 2.39, a composition in the reversibility window[13] while the latter to r = 2(1+ x) = 2.32, a composition in the flexible phase. In the data presented in Fig. 2, one can notice that the $T_g$'s of the glasses do not change upon long term aging but the sub-$T_g$ endotherm evolves. The lack of a $T_g$ shift is an important clue in modeling the sub-$T_g$ origin; the structural entity contributing to the sub-$T_g$ endotherm must form part of the network backbone since its appearance leaves $T_g$ unchanged.

#### 3. $As_xSe_{100-x}$

Our calorimetric and Raman scattering experiments reveal that long term aging effects in the As-Se binary to be rather rich with phenomenology, and we will show that these to be intimately tied to synthesis conditions and aging environment of glasses. We now present data on the four sets (**O, A, B and C**) of samples studied in this binary below.

**Trends in Glass transition temperatures ($T_g(x)$).** .

A summary of these $T_g(x)$ data on the three sets (**A, B and C**) of samples, appears in Fig. 5 $T_g(x)$ trends on samples of the original (**O**) set of samples [15] closely follow those of the rejuvenated samples (**C**). This is the case because samples of set O were aged only for a short time (3 weeks) and $T_g$s do not age that quickly. The $T_g(x)$ data in the 0% < x < 15% range (Fig. 5) show a particularly interesting behavior; at a given x aged samples display $T_g$s that exceed those of the rejuvenated ones. And as x exceeds about 15%, $T_g$'s of the three sets almost equalize. We were unable to obtain homogeneous samples at very low As concentrations in the 0 < x < 3% range, a result that is in harmony with earlier results[21]. These observations, particularly the increase of $T_g$ upon aging in the low x range, 0 < x < 15%, is an intriguing finding. We will show in section III that these trends are consistent with trigonal-Se (t-Se) and monoclinic Se (m-Se) fragments decoupling from these Se-rich glasses upon long term aging, and thus leaving behind an As-rich backbone that understandably possesses a slightly higher $T_g$.

**Trends in non-reversing enthalpy at $T_g$**. Although the magnitude of $T_g$ between the three set of samples does not change much once x exceeds 15% of As, the nature of the glass transitions undergo remarkable changes upon long term aging. That information emerges from the non-reversing heat flow scans that reveal a richness of effects. These data are summarized in Fig.3 for the slow cooled samples (**set A**), and in Fig. 4 for the melt-quenched ones (**set B**). In both figures we also include m-DSC scans of rejuvenated samples (**set C**) for comparison.

In Fig. 3, m-DSC scans of samples of **set A** at 9 compositions in the 10% < x < 40% range are presented in the sequence of increasing As content from panel (a) through (i). Each panel includes four signals- two non-reversing heat flow ones showing peaks, and two reversing heat flow signals showing rounded steps in $C_p$ near $T_g$. The continuous line traces are those of 8-year aged samples while the broken line ones are those of rejuvenated ones. A perusal of these data show the following compositional trends; (i) first, the $T_g$ deduced from the reversing heat flow appears to be nearly the same for the aged and rejuvenated samples once x > 20%, (ii) second, the non-reversing enthalpy (integrated area under the Gaussian-like profile, $\Delta H_{nr}$) is higher in the 8-year aged samples than in the rejuvenated ones at low x (< 32.5%), but the two enthalpies nearly equalize in the 32.5 % < x < 40% range. (iii) Third, a broad sub-$T_g$ endotherm is manifested near 120°C in the aged samples particularly once x > 27%, a feature quite similar to the one encountered earlier in the other selenides. These data were analyzed for the non-reversing enthalpy $\Delta H_{nr}(x)$ at $T_g$ by integrating the area under $T_g$ endotherm and sub-$T_g$ endotherm and subtracting off the exotherm observed in the cooling cycle to make the frequency correction[19]. Compositional trends in $\Delta H_{nr}(x)$ are summarized in Fig.6 as the plot of curve A. Also included in Fig. 6 are results for rejuvenated samples (**set C**) that reveal a minuscule $\Delta H_{nr}(x)$ term at all compositions. Our data reveals $\Delta H_{nr}(x)$ term for samples of **set A** to display a reversibility window that onsets near $x_1$= 33% and ends near $x_2$ = 40%. Feature (ii) above is a manifestation of the reversibility window in these 8-year aged samples (**set A**). Included in Fig. 6 for comparison are the data of Georgiev et al.[15] on melt-quenched samples aged for 3 weeks (**set O**). In comparing curves **O** with **A**, we note that the window width $\Delta x = x_2 - x_1$ of 8% for the 3 week aged samples reduces slightly to $\Delta x = 7\%$ after 8-years of aging. But perhaps more striking is the fact that upon long term aging of samples the reversibility window has sharpened a feature that we will discuss in section III.

In Fig. 4 we provide an overview of calorimetric data on melt-quenched samples aged for 8 year in laboratory environment (**set B**); the figure has 6 panels, and each panel except panels (a) and (e) shows 4 scans, 2 of the reversing- and 2 of the non-reversing- heat flow signals in the heating cycle. Panel (a) at x = 30% includes data on 3 samples aged for 3 years, 8 years and a rejuvenated sample. Panel (e) at x = 40% includes data on 4 samples one aged for 3 years, two aged for 8 years and a rejuvenated sample. In the other panels, the continuous lines refer to aged samples, while the broken line to rejuvenated ones. Starting from Fig. 4(a) at x = 30%, we observe a sub-$T_g$ endotherm and a $T_g$ endotherm for a 3 year aged sample. The sub-$T_g$ endotherm moves up in temperature and gets closer to the $T_g$ endotherm upon aging for 8 years as also observed in other glass systems. In Fig. 4 (b) at x = 32.5%, one observes the sub-$T_g$ and $T_g$ endotherm to be reasonably resolved and these terms not to increase as much upon aging as seen at lower x. In panel (c) at x = 35% we now observe for the first time evidence of a pre-$T_g$ exotherm in addition to the weak sub-$T_g$ and the $T_g$ endotherms after 8-years of aging. In panel (d) at x = 37.5% the results are quite similar to those at x = 35%. Moving on to panel (e) at x = 40%, we find one 8-year aged sample( #1) to reveal a large pre-$T_g$ exotherm while the second 8-year sample ( #2) to display <u>no</u> pre-$T_g$ exotherm. Results on a 3-year aged sample represent the intermediate case. Notice that as the strength of the pre-$T_g$ exotherm increases, that of the $T_g$ endotherm decreases, suggesting that the structural feature that contributes to the pre-$T_g$ exotherm derives from the glassy backbone. These data also reveal that samples belonging to **set B,** their heterogeneity becomes conspicuous upon long term aging. Finally in Fig. 4(f) at x = 47.5%, we observe a pre-$T_g$ exotherm followed by a weak $T_g$ endotherm in the 8-year aged sample. The range of glass compositions across which the pre-$T_g$ exotherm is manifested is approximately in the 35% < x < 60% range.

We have deduced the $\Delta H_{nr}(x)$ term for these samples of **set B** following the standard procedure. We find the $\Delta H_{nr}(x)$ term to steadily decrease as x increases from 8% to 32.5% (Fig.6). And at higher x, the $\Delta H_{nr}(x)$ term becomes nearly vanishing in the 32.5% < x < 37.5% range. But as x increases further, the $\Delta H_{nr}(x)$ term varies from sample to sample as illustrated in Fig. 4 (e) for the case at x = 40%, and for that reason the error bar associated with a measurement of $\Delta H_{nr}(x)$ becomes larger ( 0.25 cal/gm), i.e., the term becomes less well defined. Collectively, the data of Fig. 4 show that $T_g$ of glasses is unaffected by the appearance of pre-$T_g$ exotherms and sub-$T_g$ endotherms. We shall discuss the structural interpretation of these data in section III.

**X-ray Diffraction**. In Fig. 7(a) we show an XRD scan of (a) fresh melt-quenched Se glass, and in Fig. 7(c) of an 8-year aged Se glass. Glass samples in (a) and (c) were powdered just prior to the XRD measurement. In Fig. 7(e) we show the powdered 8-year aged Se glass sample re-examined 2 weeks later. During this 2 week period the sample aged in laboratory ambient environment. Also included in Fig. 7(b) and (d) are the JCPDF results[22] on trigonal (t)-Se and of monoclinic (m)-Se respectively. A perusal of these data reveal that after 8 years of aging of a Se glass (Fig. 7(c)), weak reflections of t-Se appear near 30°( 101) , 43°(102) and 45°(111). The 8 year aged Se glass after powdering and aging in the laboratory shows (Fig. 7(e)) new reflections that belong to those of m-Se (Fig.7d). Clearly, $Se_8$ rings must also form upon aging, and as the sample is powdered, i.e., as the surface/volume ratio of the glass is increased, these rings become mobile and condense to form nuclei of m-Se fairly quickly. We have re-examined the sample of Fig. 7(e) over several months thereafter, and found little change in the scan. Raman scattering data on these Se glass samples are discussed next.

**Raman scattering**. Raman scattering on samples of set B are compared to those of set C (the rejuvenated samples) in Fig. 8. Changes in Raman lineshapes between these two types of samples are small and subtle, and by plotting difference spectra (broken curve) these changes become apparent. These Raman data along with calorimetric data provide crucial insights into understanding the microscopic origin of aging. Thus, for example, a pure Se glass after 8-years aging (Fig. 8(a)) shows narrow vibrational features near 235 cm$^{-1}$, near 110 cm$^{-1}$ and near 253 cm$^{-1}$. These narrow modes are readily identified as follows: 235 cm$^{-1}$ mode is a phonon associated[23] with the symmetric stretch of helical chains of trigonal Se (t-Se), while the 110 cm$^{-1}$ and 253 cm$^{-1}$ modes are phonons associated with α-monoclinic Se. The broad mode centered near 250 cm$^{-1}$ is widely recognized[23] as a mode of disordered polymeric chains in the glassy Se.

At a finite concentration of As, i.e., at x = 6%, 8% and 10%, changes in Raman scattering upon long term aging ( Fig. 8(b), (c) and (d)) reveal the difference spectra to show three features; (i) a mode near 110 cm$^{-1}$ and (ii) a mode near 95 cm$^{-1}$ and both modes to increase in scattering strength with composition x and (iii) and the main band centered near 250 cm$^{-1}$ to red shift upon long term aging. The red shift of the main band is suggested by the sine wave like feature in the difference spectrum showing a positive excursion on the low frequency wing and a negative one on the high frequency wing of the main band near 250 cm$^{-1}$. These data show that long term aging of glasses containing a finite content of As leads to formation of Se$_8$ rings but not helical chains of Se$_n$ chains as in t-Se, as found, for example, in pure Se glass ( x = 0). At x = 20% and greater, the FT Raman lineshapes of aged samples reveal the Se$_8$ ring fraction to be nearly extinct. These Raman data when correlated with calorimetric T$_g$'s (Fig. 5) provide important clues on aging related structural changes that are discussed later in section IIIB.

At higher As concentrations (x > 35%), new features become apparent in Raman spectra of the aged samples. At x = 37.5%, the difference spectrum (dotted line in Fig 9(a)) reveals a broad feature in the 270 cm$^{-1}$ range. At x = 45%, (Fig. 9(b)) we find the difference spectrum to broaden and also shift to a lower frequency in the 235 cm$^{-1}$ range. At x = 60% (Fig. 9(c)), the difference spectrum displays many narrow modes, features that bear a striking resemblance to the vibrational modes seen in Raman scattering of c-As$_4$Se$_3$ (Fig. 9(e)). We have included Raman scattering of several crystalline compounds, such as c-$As_2Se_3$[23], c-$As_4Se_4$[24,25] and c-$As_4Se_3$[26] in Fig.9(e) for purposes of discussion of these data. A parallel result is found at x = 50% for a <u>freshly</u> quenched glass when compared to its rejuvenated counterpart in Fig. 9(d). We shall discuss these vibrational data along with pre-Tg exotherms observed in the calorimetric measurements to understand the structural manifestations of aging in these As-rich glasses in section IIID.

### III. DISCUSSION

#### A. Long term aging in pure Se glass and the nature of entropy sinks

There is general recognition that elemental Se glass consists largely of a polymeric $Se_n$ chain structure with a varying content of $Se_8$ rings[23]. The present XRD and Raman scattering data of fresh and 8-year aged Se glass (Fig. 7 and 8(a)) provide unambiguous evidence that long term aging leads to the formation of t-Se, and upon gentle crushing of the aged bulk glass also to m-Se after several weeks. The weak reflections near 30°(101), 43°(102) and 45°(111) (Fig. 7(c) are identified with formation of t-Se in the aged glass. Furthermore, the multitude of new reflections observed in Fig. 7(e), correlate well with those of α monoclinic Se (JCPDF 24-1202). In Raman scattering, modes of the t-Se phase manifest near 235 $cm^{-1}$ and those of the m-Se phase near 110 $cm^{-1}$ and 253 $cm^{-1}$ in the 8-year aged Se glass samples (Fig. 8(a)). Our best estimate of the fraction of Se atoms that crystallize upon long term aging of the bulk glass is about 2%. The estimate is based on scattering strength of the Raman modes of the glassy phase (250 $cm^{-1}$) and that of the t-Se phase (235 $cm^{-1}$) observed in the aged sample. Thus, the majority (98%) of the sample remains glassy with no change in $T_g$, but the width of $T_g$ significantly narrowed (factor of 5) and with the $\Delta H_{nr}$ term increasing from a minuscule (~ 0 cal/gm) value in fresh samples (Fig. 6) to at least an order of magnitude (about 1 cal/gm) in an 8 year aged sample. Here the width of the glass transition was measured by that of the step in the reversing heat flow associated with $T_g$, and it decreased from about 10°C for a fresh sample to about 2°C for an 8 year aged sample. These findings were made possible by recording mDSC scans at a very low scan rate of 0.3°C/min, and an example appears in Fig. 4(a) for the case of a 6% As alloyed Se glass.

How does one understand these results? The molar volume[23] of g-Se of 18.427 cm$^3$ is 12% greater than that of trigonal Se (16.385 cm$^3$) and 2.4% greater than that of α-monoclinic Se (17.99 cm$^3$). Long term aging leads the polymeric distorted Se$_n$ chains of the glassy phase to compact in a process that may be described as follows. Se atoms in distorted chains laterally diffuse (perpendicular to a local chain-axis), as the non-bonding inter-chain van der Waals interactions grow at the expense of intra-chain covalent ones. The cooperative process also leads chains to reduce in length resulting in better packing. Upon long term aging, we speculate, a given Se$_n$ chain becomes laterally correlated with about 2 or 3 neighboring chains, a reconstruction that contributes to the narrowing of $T_g$ and to the large reduction in configurational entropy of the glass as the $\Delta H_{nr}$ term increases by an order of magnitude. The ultimate fate of compaction of Se$_n$ chains, is realized by only about 2% of atoms corresponding to the case when 6 chains come laterally together to nucleate the hexagonal unit cell of t-Se[27]. The molar volume of t-Se is 16.385 cm$^3$, In such a case the chains become helical in character. A consequence of the reconstruction is the red-shift of the Se$_n$ chain mode from 250 cm$^{-1}$ in the glass to 235 cm$^{-1}$ in the t-Se phase. Thus, the physical picture of structural changes upon long term aging of a pure Se glass is that atomic groupings of the floppy chains slowly diffuse and compact as liquid Se relaxes to acquire solid glass densities. In this particular instance the compacted low entropy regions are identified with about 2 or 3 polymeric Se$_n$ chains on an average getting correlated. We shall see later that in Se-rich alloy glasses with As additive, m-Se fragments are nucleated in the glass upon long term aging. These aging results on elemental Se are not unexpected given that its $T_g$ of 40°C, is quite close to room temperature where our samples were aged.

As a final comment we would like to emphasize that Raman scattering data on fresh and aged Se glass were obtained by using a low power density 1064 nm radiation in an FT Raman set up. It is well known that glassy Se can be easily photocrystallized[28,29] in Raman scattering experiments utilizing visible (632 nm) exciting radiation. It is also known that Se glass can be crystallized by crushing glass lumps. The results described in this work are not related to these effects. The segregation of the crystalline Se polymorphs upon long term aging of Se glass will have an important bearing in understanding long term aging effects of Se-rich As-Se glasses as we discuss next.

### B. Microscopic origin of $T_g$ increase upon long term aging in Se-rich glasses.

Historically, discussions of the glass transition have emphasized its kinetic nature[30-32] and the magnitude of $T_g$ related to rate of undercooling or melt quenching. It is generally believed that α-relaxation processes slow down, as relaxation time ($\tau_\alpha$) diverges to become infinitely large as T approaches $T_o$, the ideal glass transition temperature with $1/\tau_\alpha$ following a Vogel - Fulcher -Tamman variation[33]. Typical structural relaxation times near $T_g$ are in the 100 sec range. In our experiments these kinetic effects can only play a minor role because not only all glass samples were synthesized at the same cooling rate but also because our measurements of $T_g$ in mDSC scans have used very slow scan rates ( 3°C/min) leading typically to a scan time of 400 seconds across a 20°C wide glass transition endotherm. On the other hand, changes in $T_g$ brought about by chemical alloying group IV and/or group V additives in a base Se glass far exceed those brought about by quench rates and scan rates. Variations in $T_g$ brought about by chemical alloying, it has been shown[15,34], can be quantitatively understood in terms of Stochastic Agglomeration Theory with few adjustable parameters. In the latter approach, the magnitude of $T_g$ is found to be a faithful representation of glass network connectivity[35,36]. These ideas connecting *glass molecular structure* to the nature of glass transitions including the magnitude of $T_g$ and its non-ergodicity is a more recent development in the field[37,38].

The increase of $T_g$ upon long term aging of Se-rich binary As-Se glasses (Fig. 5), forms part of fairly general observation in the field. Other groups[21,30,39] have also reported such an observation in Se-rich glasses containing less than 10 at. % of group IV and/or group V additives. In these samples long term room temperature aging will lead to demixing of t-Se and /or m-Se fragments from the network backbone leaving it Se-deficient. In a pure Se glass, decoupling of t-Se and m-Se impurity phases from the base glass network will not affect its $T_g$. However, in a binary ($Pn_xSe_{100-x}$, $Tt_ySe_{1-y}$)) or ternary ($Pn_xTt_ySe_{100-x-y}$) Se-rich alloy glass, wherein additives (Pn= Pnictide, Tt= Tathogen) crosslink the backbone, one can expect a further increase of $T_g$ to occur as a consequence of aging as the alloyed network becomes Se depleted. The $T_g$ increase due to aging can be expected to depend linearly on the fraction of demixed t-Se and/or m-Se from the alloyed glass.

As the alloying concentration of the additives (Pn, Tt) increases, one can expect an onset of a *saturation* of the $T_g$ increase with aging to occur. As $T_g$'s climb much above room temperature upon progressive alloying, the demixing of t-Se will slow down for obvious kinetic reason. It will slow down also because of a topological factor; finding $Se_n$ chain segments of any length that survive, let alone come together and compact will become less likely upon progressive alloying. These considerations will not hinder $Se_8$ rings to decouple from the backbone however. Our results support such a physical picture of aging at low As content. The Raman scattering data of Fig. 8(b), (c) and (d) reveal growth of modes near 110 cm$^{-1}$ and 95 cm$^{-1}$ ascribed to Se$_8$ rings. Nagata et al.[40] have shown α-monoclinic Se to have an IR active E$_1$ mode near 97 cm$^{-1}$ that appears to be Raman active as well. The decoupling of Se$_8$ rings upon long term aging leaves the alloyed As-Se networks progressively As-rich upon aging, and will lead T$_g$ to increase as reflected in Fig. 5 in the narrow interval 0 < x < 15%. At higher x (> 20%), Raman scattering provides no evidence of either t-Se or m-Se impurity phases decoupling from glasses. In fact, Raman scattering of aged and rejuvenated samples are virtually identical at x = 20% .At these higher As concentrations there is no longer enough free Se left to nucleate either t-Se or m-Se fragments in glasses. These ideas provide a natural basis to understanding the higher slope of T$_g$ with x near x ~10% (Fig. 5) upon long term aging of slow cooled and melt quenched Se-rich glasses.

Selenium rich glasses when alloyed with increasing concentration of Ge show the broad Se-chain band (~250 cm$^{-1}$ ) to steadily *blue-shift*.[41] The *red-shift* of the main Se-chain band in the present aging experiments on As-Se glasses is, therefore, particularly curious. Raman lineshapes of fresh binary As$_x$Se$_{100-x}$ glasses, have also been examined by us systematically with As content over a wide range,[17] and in the low x range ( ~25%) , we find growth of modes near 235 cm$^{-1}$ and 220 cm$^{-1}$ ascribed respectively to normal modes of pyramidal (AsSe$_3$) and quasi-tetrahedral (Se=As(Se$_{1/2}$)$_3$)) local structures formed as As cross-links chains of of Se$_n$. And as glasses age and backbones become Se-rich, one expects these modes to also grow at the expense of the Se$_n$ chain modes. We believe the red-shift of the main chain mode in the aging experiments is due to the presence of these modes, which overwhelms blue-shift of the stiffened

Se$_n$ chains as they shorten, and account for the third feature observed in the spectra of Fig. 8(b)-(d) that we alluded to earlier.

Why is the T$_g$ increase with x in As-Se glasses upon long term aging greater in *melt-quenched* glasses than in *slow cooled* ones (Fig. 5)? The answer rests in melt-quenched glasses being less homogeneous than slow cooled ones; their networks possess larger spatial fluctuations in atomic concentrations from the average than in slow cooled ones. One can expect more Se-rich and As-rich regions to form in melt-quenched samples (**set B**) than in slow cooled ones (**set A**). The Se-rich regions assist in decoupling of m-Se, and lead, indirectly, to a further increase of T$_g$. The As-rich regions directly lead to a T$_g$ increase because these regions are more cross-linked. Although these T$_g$ shifts are small, they are systematic, and the narrowing of glass transitions upon long term aging, permits their measurement to an accuracy of less than 0.5°C in these cases. Since the scan rates used are extremely small (0.3°C/min), there are virtually no scan rate related shifts in T$_g$. And as the As concentration exceeds 15%, differences in T$_g$ between melt-quenched (T$_g^{mq}$) and slow cooled (T$_g^{sc}$) samples with rejuvenated (T$_g^{rej}$) ones, vanish. We expect the pattern to be observed in other binary and ternary selenide glasses as well. The structural heterogeneity of melt quenched As-Se glasses, particularly the As-rich regions will also have a bearing on the observation of the pre-T$_g$ exotherms, an issue we discuss in section IIID.

### C. Sub-T$_g$ endotherms and secondary relaxation in selenide glasses

A significant finding of the present work is the observation of sub-T$_g$ endotherms upon long term aging of selenide glasses. These endotherms are observed as peaks, typically as wide as T$_g$ endotherms, but shifted 10°C to 70°C below T$_g$. They are observed in binary and ternary alloys that we have investigated (Fig. 2, 3, 4) here. Sub-T$_g$ endotherms display no specific heat jump in the reversing heat flow, a behavior that contrasts with that of the glass transition endotherm. For that reason, sub-$T_g$ endotherms are better understood as activated processes, rather than secondary glass transitions. What kinds of processes are involved in sub-$T_g$ endotherms?

We identify sub-$T_g$ endotherms with parts of the backbone that have compacted. These regions, usually are mildly cross-linked chains of Se (with As and Ge atoms) that slowly diffuse with respect to each other to lower molar volume of a glass globally upon long term aging. The physical processes contributing to the sub-$T_g$ endotherms are essentially the same ones that contribute to compaction of Se-chains in the glassy Se discussed in section IIIA. The eventual fate of the compacted regions is <u>not</u> a crystalline product that demixes from the backbone, but it is a glassy product that remains part of the network. These *compacted regions* of the present chalcogenide glasses studied must possess lower entropy and are viewed as traps in the Scher-Lax-Phillips model[7] of stretched exponential relaxation of a glassy network. Our experiments also reveal that the sub-$T_g$ endotherms steadily up-shift in temperature upon long term aging suggesting that the underlying activation energy must increase with aging time. Such an increase may be tied to the concentration of the crosslinking atoms (As, Ge, P) increasing in the *compacted regions*. Once an aged glass is cycled through $T_g$, the sub-$T_g$ endotherms completely vanish as *compacted regions* melt and coalesce with the *non-compacted* ones essentially restoring the larger free volume characteristic of the fresh glass. And as a fresh glass is aged, sub-$T_g$ endotherms steadily reappear, thus showing reproducibility of the aging process. The view that sub-$T_g$ endotherms represent compacted regions of a flexible glass upon long term aging finds support in a recent enthalpy lansdscape model of Se developed by Mauro and Loucks.[42]

In a study several years ago we noted observing sub-$T_g$ endotherms in binary $P_xSe_{100-x}$ glasses (x = 37%) when they are synthesized by melt quenching but not when they are slow cooled[43]. The observation is related to the disproportionation of quasi-tetrahedral $Se=P(Se_{1/2})_3$ units present in melts into pyramidal $P(Se_{1/2})_3$ units and $Se_n$-rich phase for which evidence emerged independently from $^{31}P$ NMR experiments[44]. The sub-$T_g$ endotherm (= 77°C) observed in m-DSC experiments on a $P_{37}Se_{63}$ glass was identified with the presence of $Se_n$-rich regions formed in such melts. The sub-$T_g$ endotherm here, of course, is manifested in a process that is quite different from the aging experiments described above, but the common feature remains that it is identified with a flexible part of the network in both cases, an important diagnostic feature.

In polymers sub-$T_g$ transitions are observed in DSC, TMA and DMA experiments and are generally associated with localized movement of main chains or large side chains, and alter toughness of these materials.[45] These have been described as secondary relaxation by Johari[46] and Goldstein[47] as contributing to aging of glasses by compaction of networks to lower entropy and enthalpy leading to corresponding increases in refractive index[48] and shear moduli[45]. These physical effects are quite similar to the ones described in this work on chalcogenides.

### D. Pre-$T_g$ exotherms and NSPS of glassy As-Se networks

An important finding of the present work is the observation of *pre-$T_g$ exotherms* in melt-quenched As-Se glasses aged in the laboratory environment (**set B**) but not in the slow cooled samples of **set A** that were aged in total darkness. These exotherms are observed typically in the 35% < x < 60% range (Fig. 4). What is their microscopic origin?

It is known that As-As bonds are first manifested in $As_xS_{100-x}$ bulk glasses[49] when the As content exceeds 35% of As. A similar behavior can be expected in As-Se glasses as well. These homopolar bonds can either form part of polymeric ethylene-like $As_2(Se_{1/2})_4$ polymeric chains[24,50], as for example found in a bulk $As_{50}Se_{50}$ glass, or they can form part of monomers composed of $As_4Se_4$ and $As_4Se_3$ molecular cages. A parallel circumstance occurs in binary $P_xSe_{100-x}$ glasses wherein polymeric chains made up of ethylene-like $P_2(Se_{1/2})_4$ units are observed once x exceeds 25%[43], and at higher P-concentration, x >50%, $P_4Se_3$ cages are found to rapidly decouple from the backbone to preclude bulk glass formation at x = 57%. It is also known that bulk $As_{50}Se_{50}$ glass, in a finely powdered form, can be completely transformed[24] to a molecular crystalline solid composed of $As_4Se_4$ molecules, the crystalline counterpart of $As_4S_4$ also known as Realgar, by a low-temperature (150°C) thermal annealing as shown recently[24]. The same crystalline phase can be formed by exposing amorphous $As_{50}Se_{50}$ thin-films to visible light as reported by Kolobov and Elliott[25]. Experiments also show that such photo-crystallized films can be re-amorphized by light exposure. The reversibility of $As_{50}Se_{50}$ from a polymeric glass structure into a crystalline molecular ($As_4Se_4$) solid, is at the base of structural changes underlying long term aging of the present As-rich glasses in a natural ambient environment.

The origin of the pre-$T_g$ exotherm observed in the m-DSC experiments can now be described. As an aged glass is heated above room temperature these molecules begin to diffuse as the molecules are weakly bonded ( by van der Waals forces) to the base network. And as the T exceeds about 120°C, these molecules begin to condense forming nuclei of the molecular $As_4Se_4$ crystal. Since the free energy of this condensed molecular phase is lower than that of the phase of these molecules at room temperature, one expects heat to be released as a glass is heated in the 110°C < T < 160°C. We have already noted that c-$As_4Se_4$ [24] can be grown by merely heating $As_{50}Se_{50}$ glass at 150°C for 4 days. And as T increases further to $T_g$, $As_4Se_4$ molecules remix with the backbone to form a bulk glass with the same $T_g$ as that of the rejuvenated glass. A perusal of the mDSC scans of samples at x = 37.5%, 40% and 47.5%, illustrated in Fig. 4 (d), (e) and (f), show not only the presence of pre-$T_g$ exotherms (in the non-reversing heat flows) but also $T_g$'s of the aged and rejuvenated glasses to be nearly identical (in the reversing heat flow steps). In Raman scattering data of Fig.9 (a), (b) and (c)), we compare lineshapes of these As-rich aged glasses with their and rejuvenated counterparts. For reference purposes, we have included in Fig. 9(e), Raman scattering of various crystalline phases[24,26], c-$As_2Se_3$, c-$As_4Se_4$ and c-$As_4Se_3$. Of interest is the Raman lineshape difference between the aged and rejuvenated sample in a glass at x = 37.5% shown as a dotted curve in Fig. 9(a); note that an excess of vibrational density of states is observed near 260 $cm^{-1}$, with weaker features near 200 $cm^{-1}$ and 150 $cm^{-1}$ upon aging. These features are quite similar to those observed in Raman scattering of c-$As_4Se_4$ shown in Fig. 9(e). At x = 45% (panel b), the excess vibrational density of states between the aged and rejuvenated samples now broadens and shifts to lower frequencies to include a mode near 235 $cm^{-1}$. The latter vibrational mode correlates well with the strongly excited phonon observed in Raman scattering of c-$As_4Se_3$ (Fig. 9(e)). At the glass composition x = 60%, the Raman difference signal (Fig. 9(c)) now reveals many features also seen in c-$As_4Se_3$. At x = 60%, partial demixing of the base network into isolated molecular units is a feature of even freshly prepared glasses as noted earlier in inelastic neutron scattering results of Effey and Cappelletti[51]. One can see here a distinct pattern to emerge: as the As content of the glasses increases the nature of molecular species decoupling from the backbone upon long term aging changes from being predominantly $As_4Se_4$ at x = 37.5%, to a mix of both $As_4Se_4$ and $As_4Se_3$ molecules at x = 45%, and to becoming predominantly $As_4Se_3$ at the highest As concentration of

x = 60%. The experimental evidence suggests that As-rich (x > 35%) melt-quenched As-Se glasses, when aged at laboratory ambient environment, steadily disproportionate with $As_4Se_4$ and $As_4Se_3$ molecules demixing from backbones. The underlying process is believed to be light assisted in samples of **set B**, since such demixing is not observed in samples that are aged in total darkness (**set A**). The correlation of Raman scattering data (Fig 9) with mDSC data (Fig 4) reveals high sensitivity of the calorimetric measurements to NSPS ; even less than 5% of As atoms segregated in the form of molecules from the base network are apparently enough to show the exothermic feature.

The pre-$T_g$ exotherm observed in the present As-Se glasses is reminiscent of another calorimetric study[52] on binary $As_xS_{100-x}$ glasses, where one also observed an exotherm but this time as a precursor to the sulfur ring to chain ($S_8 \rightarrow S_n$) polymerization transition near $T = T_\lambda \sim$ 125°C. In these sulfides, at low As content (x < 10%), one has a glass network that largely consists of $S_8$ rings that demix from the backbone composed of As cross-linked $S_n$ chains. Upon heating these glasses past their $T_g$, one encounters an exotherm as a precursor to the $T_\lambda$ transition. The exotherm is observed, typically in the 60°C < T < 110°C range. It is identified with $S_8$ rings becoming mobile as T > 60°C, and coalescing to lower the free energy of the system ( and expelling heat ) by nucleating a $S_8$ ring phase prior to these rings opening and polymerizing to form chains as $T > T_\lambda \sim$ 125°C. The structural interpretation of exotherms in both instances can be traced to monomeric species that becomes mobile at an elevated temperature prior to their breaking up as $T_g$ is approached as in the present As-Se glasses, or to a sulfur polymerization transition as in the case of the As-S glasses.

Finally, it is useful to mention that pre-$T_g$ exotherms can also arise due to an entirely different physical mechanism; stress frozen in on account of a fast quench of melts as observed for example in Te-As-Se fibers[53]. Surfaces of fibers undergo a faster thermal quench than fiber interior with the consequence that they are trapped at lower mass densities. And since their $T_g$s are close to room temperature, long term aging displays pre-$T_g$ exotherms as the frozen stress relaxes. Stress can also be frozen upon a fast quench of bulk glasses, and upon heating a pre-$T_g$ exotherm can be observed as is the case of $AgPO_3$.[54] . Aging in ceramic oxides such as clays display profound hydroxylations effects[55], a feature also noted in $AgPO_3$ [54], and a behavior that contrasts to the one noted here in chalcogenides.

### E. Comparison with the m-DSC results of Golovchak et al.

Golovchak et al.[14,56] have recently reported m-DSC results on furnace cooled $As_xSe_{100-x}$ glass samples that were aged in hermetically sealed plastic bags, which were kept in the dark for 22 years. Their measured non-reversing enthalpy, $\Delta H_{nr}(x)$, for the 9 sample compositions are the inverted triangles (curve G)) plotted in Fig. 6(a). The filled (green) circles (curve B) are $\Delta H_{nr}(x)$ for samples in our set B. The much earlier measurements of Georgiev et al.[15] on these samples, aged for only 3 weeks, are given by the filled (red) triangles as curve O in that figure. The $\Delta H_{nr}(x)$ data on samples of set A (filled square, curve A) and for rejuvenated samples (open circles, curve C are plotted in Fig.6 (b).

Golovchak et al.[14] have also provided non-reversing heat flow scans of their aged samples along with those of the rejuvenated ones in Fig. 3 of Ref. [14]. It is instructive to compare their data with those on present samples of sets A and B provided in Fig. 3 and 4 respectively. Several observations can be made. (i) $T_g(x)$ trends of the rejuvenated samples inferred from the peaks in the non-reversing heat flow in Fig. 3 of Ref. [14] track well the $T_g(x)$ trends on the present rejuvenated samples (filled circles) shown in Fig. 5. These data suggest that glass compositions (As content x) synthesized by both groups are in reasonable agreement with each other.(ii) For samples at x = 40% and x =50% (Fig. 3 of Ref. [14]), Golovchak et al.[14] observe evidence of *pre-$T_g$ exotherms*, similar to the ones seen in samples of set B across the range of compositions, 40% < x < 60% (Fig. 4). (iii) At x = 10% , they observe $T_g$ of their aged sample to be about 17°C higher than the rejuvenated one. The corresponding $T_g$ shift of our melt quenched sample (**set B**) is 12°C, and for the slow cooled one (**set A**) is 5°C. Taken together, these data show that aging behavior of samples studied by Golovchak et al. [14] is similar to that of our samples of **set B**.

In samples of **set A** which were aged in hermetically sealed Al pans (TA Instruments) in total darkness, we find no evidence of *pre-$T_g$ exotherms* at any of the compositions examined (Fig. 3). These were samples sealed in Al pans by D. Georgiev et al.[15] in the year 2000. The absence of the *pre-$T_g$ exotherm* in these samples permits a straightforward analysis to extract the frequency corrected non-reversing enthalpy ($\Delta H_{nr}(x)$). These data reveal the reversibility window to be in the 33% < x < 40% range (Fig. 6(b)). We believe results on samples of set A represent the *intrinsic* aging behavior of these *isolated* glasses.

We turn to exotherms. These were not commented upon by Golovchak et al.[14] but we now believe them to be central to an understanding of the structural changes taking place in aged samples. They lead to an apparent reduction in $\Delta H_{nr}(x)$. This can be clearly seen in the m-DSC scans on three g-$As_{40}Se_{60}$ samples shown in Fig. 4(e). One of these samples labeled #1 shows the largest *pre-$T_g$ exotherm*, but also the smallest $\Delta H_{nr}(x)$ term at $T_g$. Sample #2 shows the smallest *pre-$T_g$ exotherm* but also the largest $\Delta H_{nr}(x)$ term at $T_g$. The third sample aged for 3 years represents the intermediate case. Clearly, once a *pre-$T_g$ exotherm* is manifested, the $\Delta H_{nr}(x)$ term in general <u>decreases</u> in strength.

Nanoscale phase separation (NSPS) in network glasses[57], in general, leads to demixing of backbones, or loss in network connectivity which is reflected in $\Delta H_{nr}(x)$ term. In fresh $Ge_xP_xSe_{1-2x}$ glasses, S. Chakravarty et al.[11], observed a satellite window in the $\Delta H_{nr}(x)$ term near x = 22%, corresponding to *r* = 2.66, due to demixing of some $P_4Se_3$ molecules in these glasses. NSPS effects are also pronounced in binary $As_xSe_{100-x}$ glasses particularly once x > 40% as reflected in $T_g(x)$ decreasing with increasing x (Fig. 5). At a glass composition x = 60%, we have already mentioned inelastic neutron scattering studies of Effey and Cappelletti[51], who have shown sharp modes characteristic of intra-molecular vibrations of $As_4Se_4$ molecules to be present. These molecules demix from the backbone, resulting in a loss of network connectivity, which is reflected in the growth of floppy modes (near 5 meV) and a low $T_g$, characteristic features of a flexible glass even though *r* = 2.60!! At x = 50%, we have already alluded to the reversibility of this composition from a polymeric glass structure into a crystalline $As_4Se_4$ molecular solid and vice-versa by the action of heat or by light [25].

Returning to the mDSC scans of Fig 4(e), as T increases to $T_g$, a structural reconstruction of the demixed molecules with the backbone must take place, and such reconstruction must erase in good part the aging induced increase of the non-reversing enthalpy of the backbone stored at room temperature. The vanishing value of $\Delta H_{nr}(x)$ term found by Golovchak et al.[14] and by us for samples of set B at x > 40% is the result of NSPS of backbones. These results elegantly show that long term aging of As-rich glasses leads NSPS effects to be extended from to As concentrations of x = 60% to 40%. These results are characteristic of light modified aging of these As-rich glasses, i.e., aging caused by *extrinsic* effects.

The similarity of $\Delta H_{nr}(x)$ data between curve B (present work) and curve G (Golovchak et al. Ref. [14]) in Fig. 6(a) extends to Se-rich glasses as well. For example, the $\Delta H_{nr}(x)$ term of samples at x = 0, 10%, 20% and 30% between the two sets are similar suggesting that the increase of the non-reversing enthalpy due to aging between 8 years (curve B) and 22 years (curve G) is minimal, i.e., the aging behavior has nearly saturated after 8 years. In their investigations, Golovchak et al.[14] examined a glass composition at 30% and one at 40% but none in between. If they had made measurements in the IP composition range, 30% < x < 40%, we feel confident that they too would have observed the reversibility window as we did in samples of **set B**.

### F. Long- term aging of the reversibility window in $As_xSe_{100-x}$ glasses

The results on the non-reversing enthalpy at $T_g$ (Fig.6) permit commenting on long term aging of the reversibility window (RW) in binary As-Se glasses. We begin by recognizing that the $\Delta H_{nr}(x)$ trends on the rejuvenated samples provide a baseline (curve C Fig. 6(b)) for the aging experiments, essentially setting the waiting time clock, $t_w = 0$. The minuscule $\Delta H_{nr}(x)$ term for rejuvenated glasses suggests that their configurational entropies are liquid-like in the initial stages ($t_w \sim 0$). Georgiev et al.[15] have examined melt-quenched samples allowed to relax at room temperature for about 3 weeks prior to initiating a heating cycle in an m-DSC experiment. For these samples, the $\Delta H_{nr}(x)$ term has steadily evolved, it increased at low x and at high x, but not at intermediate x leading to the RW (curve O in Fig. 6(a)) in the 29% < x < 37% range with a width $\Delta x = 8\%$. These samples were brought to room temperature after $T_g$ cycling at a slow cooling rate of 3°C/min, and aged for 8 years in the same hermetically sealed Al pans used in the year 2000. Curve A of Fig. 6(b) reveals trends in the $\Delta H_{nr}(x)$ term after 8 years of aging, to display a RW that is now sharper and somewhat narrower (33% < x < 40%;

Δx = 7%), with a centroid at 36.5%, slightly up-shifted from the centroid (33%) of samples of **set O**. In Intermediate phases networks acquire the new capability to rewire and expel stressed or redundant bonds, and thereby lower their free energy. The sharpening of the lower end of the window (Fig. 6(b)) is viewed to be a consequence of such reconnecting. These aging results support the notion of Intermediate phases as self-organized phases of glasses. These findings directly contradict the premise advanced in ref [14] that reversibility windows cannot be established upon aging for short duration (weeks) and that upon long term (years) aging the window in As-Se glasses shifts to compositions x >40%. The aging pattern of the reversibility window observed in the present As-Se binary forms part of the general pattern noted in other chalcogenides[10,11,16,17]. Glassy networks in the elastically flexible and stressed-rigid phases generally age but those in Intermediate phases almost do not as demonstrated for the Ge-P-Se[11], Ge-As-Se[16] and the Ge-Se[10,17] systems.

The RW onset near x = 33% or $r$ = 2.33 < 2.40 in the present As-Se binary is similar to the case of other group V chalcogenides including P-Se[43], P-S[58], As-S[52], Ge-P-Se[11], Ge-As-Se[16]. The window onset at $r$ < 2.40 (here 2.40 corresponds to the the mean-field rigidity onset value[59]) is the consequence of stress-free networks formed by two isostatic building blocks[58] present in these networks, a pyramidally (PYR) coordinated ($Pn(Ch_{1/2})_3$ and a quasi-tetrahedrally (QT) coordinated ($Ch=Pn(Ch_{1/2})_3$ group V atom (Pn). Here Pn designates a Pnictide and Ch a Chalcogen atom. Constraint counting algorithms[60] show these two local structures to be isostatic[52] even though their mean coordination number $r$ = 2.28 and 2.40 respectively. The existence of quasi-tetrahedral $S=As(S_{1/2})_3$ local structure in binary As-S glasses was inferred[52] recently from analyzing Raman vibrational density of states using first principles cluster calculations. The case for the quasi-tetrahedral $Se=As(Se_{1/2})_3$ local structure in binary As-Se glasses was suggested from first principles MD calculations[61]. The experimental evidence for Se=As(Se$_{1/2}$)$_3$ local structure continues to be indirect, and it comes from the $T_g(x)$ trends in rejuvenated glasses that reveal a slope[15] much too small to be accounted by PYR units as the sole building block of these glasses at low x (< 15%). The conclusion is also supported by compositional trends in As-Se liquid fragilies[37] and compositional trends in thermal conductivity,[62] which reveal a global minimum in the RW. These results may be contrasted to those of group IV chalcogenides, such as Ge-Se[41,63] and Si-Se[64,65] glasses, which reveal the RWs to be in the 2.40 < r < 2.52 range. In these glasses the corresponding isostatic local structural units include corner sharing - and edge sharing - tetrahedra[59,66-69] that possess a mean coordination number that equals 2.40 and 2.67 respectively, primarily responsible for a shift of the RWs to $r$ > 2.40, the mean-field rigidity transition value[59].

## IV. CONCLUSIONS

Long term aging extending from months to several years is studied on several families of chalcogenide glasses (Ge-Se, As-Se, Ge-P-Se, Ge-As-Se) using modulated DSC and FT-Raman scattering. Two sets of As$_x$Se$_{100-x}$ samples aged for 8 years were compared, **set A** consisted of slow cooled samples aged in hermetically sealed Al pans in total darkness, and **set B** consisted of melt quenched samples aged in laboratory environment.

Our results show all samples display a sub-$T_g$ endotherm typically 10°C to 70°C below $T_g$. In the As-Se binary, samples of **set B**, in addition, display a pre-$T_g$ exotherm in the As range of 35% < x < 60%, a feature not observed in samples of **set A**. We identify sub-$T_g$ endotherms with a progressive compaction of flexible networks upon long term aging. The pre-$T_g$ exotherm result due to Nano Scale Phase Separation (NSPS), i.e., demixing of As$_4$Se$_4$ and As$_4$Se$_3$ molecules from the backbone. NSPS effects are accentuated in glassy systems where there is a propensity for polymeric networks to disproportionate into small molecular units, as in the case of group V chalcogenides.

The reversibility window in As-Se glasses of **set A** is in the 33% < x < 40% range after 8 years of aging; it is sharper and somewhat narrower than the window observed after 3 weeks of aging (29% < x < 37%) reported earlier by Georgiev et al. [15] . The sharpening and narrowing of the reversibility window upon long term aging in As-Se binary glasses is viewed as the slow 'self-organizing' stress relaxation of the phases outside the Intermediate phase, which itself is stress-free and displays little aging. These findings contradict the premise advanced by Golovchak et al.[14] that reversibility windows cannot be established upon aging for short duration (weeks), and that upon long term (years) aging the window onset in As-Se glasses shifts to the stoichiometric composition (x = 40%). The vanishing non-reversing enthalpy in the 40% < x < 55% range , we have seen to represent NSPS of the glasses as also found in our samples of set B in the present work.

**Acknowledgements**

We have benefitted from discussions with Dr. Bernard Goodman, Dr. Darl McDaniel, Jacob Watchman, Dr. Matthieu Micoulaut, Dr. John Mauro, Len Thomas and Steve Hall. The present contribution represents in part the Ph.D. Dissertation work of Ping Chen at University of Cincinnati. The data on the Ge-As-Se ternary was taken from the Ph.D. Thesis work of Tao Qu. This work is supported in part by the NSF grant DMR- 04-56472 and DMR-08-53957.

**Figure captions**

**FIG. 1.** Handling of As-Se binary glass samples schematically represented as a chain of custody serving to define the four sets of samples, **set A**, **set B** and **set C** and **set** O used in the present work. **Set O** (original) refers to samples used by Georgiev et al. in the year 2000 (ref. 15) and were aged in quartz tubes for 3 weeks after a water quench. **Set A** refers to the actual mDSC samples of **set O** aged for 8 years in Al pans. **Set B** refers to samples of **set O** aged for 8 years in plastic vials with push caps at laboratory ambient environment. **Set C** refers to samples that were rejuvenated after aging. Data on samples of **set G** is taken from Golovchak et al. ref. [14].

**FIG. 2.** MDSC scans of $Ge_{16}Se_{84}$ glass shown in (a) and (b), and of $Ge_{13}As_{13}Se_{74}$ glass shown in (c) and (d); these data show evolution of the sub $T_g$ endotherm with waiting time. Note in the flexible glass $Ge_{16}Se_{84}$ the sub-Tg endotherm moves up in temperature with aging but in the intermediate phase glass $Ge_{13}As_{13}Se_{74}$ there is little or no movement of the sub-$T_g$ feature.

**FIG. 3.** MDSC scans of $As_xSe_{100-x}$ samples of set A (slow cooled and aged in dark for 8-years) compared to those of rejuvenated samples. Data at 9 indicated compositions are shown with increasing As content in the sequence (a) through (i). Each panel includes reversing and non-reversing heat flow scans with solid lines used for slow-cooled samples and dashed lines used for rejuvenated ones.

**FIG. 4.** MDSC scans of samples of set B (water-quenched and aged at laboratory environment) at 6 compositions shown with increasing As content in the sequence (a) through (f). Each panel shows reversing and non-reversing heat scans for aged (solid, dash dot, dash dot dot) and rejuvenated (dash) $As_xSe_{100-x}$ glasses. Notice the pre-$T_g$ exotherm is observed in the aged samples once x > 35%. Its presence makes an unambiguous analysis of the non-reversing enthalpy at $T_g$ difficult at best (see text).

**FIG. 5** (Color online). Glass transition temperatures of binary $As_xSe_{100-x}$ glasses determined from 3 sets of MDSC scans: 8-year aged melt-quenched (triangle, $\Delta$), 8-year aged slow-cooled (open circle, ○) and rejuvenated (filled circle, ●). See text for details.

**FIG. 6.** (Color online). Compositional trends in non-reversing enthalpy $\Delta H_{nr}(x)$ for (a) water-quenched samples of set O(filled triangle ▲, from D.Georgiev et al.ref. 15) and set B (filled circle ●, present work) and samples of Golovchak et al. ref 14 shown as set G (inverted filled triangle ▼). In panel (b), corresponding data for samples of set A (filled square ■) and rejuvenated samples (open circles ○) is included. Notice the close similarity of the data between samples of set G and set B.

**FIG. 7.** XRD scans of (a) fresh Se glass (c) an 8- year aged Se glass and (e) sample in (c) aged in the laboratory environments for two weeks. Also included in figure are JCPDF reflections for t-Se in (b) and α-monoclinic Se in (d).

**FIG. 8.** (Color online) Raman scattering of (a) Se, (b) $As_6Se_{94}$, (c) $As_8Se_{92}$ and (d) $As_{10}Se_{90}$ glasses. At each composition, we show 8-yr aged (solid, blue),fresh or rejuvenated glass (dash, black), and difference spectra (dotted line, red) are shown. Notice the t-Se peak appears near 2357 cm$^{-1}$ in the aged sample. In the As alloyed glasses $Se_8$ ring modes appear upon long term aging.

**FIG. 9.** (Color online) Raman scattering of (a) $As_{37.5}Se_{62.5}$, (b) $As_{45}Se_{55}$, (c) $As_{60}Se_{40}$ ,(d) $As_{50}Se_{50}$, and (e) crystalline $As_2Se_3$, $As_4Se_4$, and $As_4Se_3$. In panels (a)-(c), we show data on 8-year aged samples (solid, blue), rejuvenated samples (dash, black), and difference spectra (dotted line, red) are shown. In (d), we compare Raman spectra of water-quenched fresh glass (solid,

blue) with the rejuvenated (dash, black) sample; the difference spectrum shows modes similar to those found in c-$As_4Se_3$.

*Present address: Dept. of Electrical & Computer Engineering, Boise State University, 1910 University Dr. Boise, ID 83725-2075, USA

**References**

[1]  Phillips J C 1996 *Rep. Prog. Phys.* **59** 1133
[2]  Kohlrausch R 1847 *Poggendorf's Ann. Physik Chem.* **72** 353
[3]  Denny R A, Reichman D R and Bouchaud J-P 2003 *Phys. Rev. Lett.* **90** 025503
[4]  Sturman B, Podivilov E and Gorkunov M 2003 *Phys. Rev. Lett.* **91** 176602
[5]  Xia X and Wolynes P G 2001 *Phys. Rev. Lett.* **86** 5526
[6]  Palmer R G, Stein D L, Abrahams E and Anderson P W 1984 *Phys. Rev. Lett.* **53** 958
[7]  Phillips J C 2009 *arXiv:0903.1067v1*
[8]  Boolchand P, Georgiev D G and Goodman B 2001 *J. Optoelectron. Adv. Mater.* **3** 703
[9]  Boolchand P, Lucovsky G, Phillips J C and Thorpe M F 2005 *Phil. Mag.* **85** 3823
[10] Wang F, Mamedov S, Boolchand P, Goodman B and Chandrasekhar M 2005 *Phys. Rev. B* **71** 174201
[11] Chakravarty S, Georgiev D G, Boolchand P and Micoulaut M 2005 *J. Phys. Condens. Matter* **17** L1
[12] Boolchand P, Chen P, Jin M, Goodman B and Bresser W J 2007 *Physica B* **389** 18
[13] Qu T, Georgiev D G, Boolchand P and Micoulaut M 2003 The intermediate phase in ternary $Ge_xAs_xSe_{1-2x}$ glasses in *Supercooled Liquids, Glass Transition and Bulk Metallic Glasses* (Materials Research Society) Vol. 754
[14] Golovchak R, Jain H, Shpotyuk O, Kozdras A, Saiter A and Saiter J M 2008 *Phys. Rev. B* **78** 014202
[15] Georgiev D G, Boolchand P and Micoulaut M 2000 *Phys. Rev. B* **62** R9228
[16] Qu T 2004 *Ph.D. Thesis* University of Cincinnati
[17] Chen P 2009 *Ph.D. Thesis* University of Cincinnati
[18] Massalski T B and ASM International. 1990 *Binary alloy phase diagrams* (Materials Park, Ohio: ASM International)
[19] Thomas L C 2006 *Modulated DSC Technology (MSDC-2006)* (New Castle, DE: T.A. Instruments, Inc)
[20] Verdonck E, Schaap K and Thomas L C 1999 *Int. J. Pharm.* **192** 3
[21] Tonchev D and Kasap S O 2002 *Mater. Sci. Eng. A* **328** 62
[22]  *Materials Data, Inc., Jade 5.0*
[23] Zallen R and Lucovsky G 1974 The Interaction with Light of Phonons in Selenium in *Selenium* (New York: Van Nostrand Reinhold)
[24] Wachtman J L 2009 *MS Thesis* University of Cincinnati
[25] Kolobov A V and Elliott S R 1995 *J. Non-Cryst. Solids* **189** 297
[26] Bues W, Somer M and Brockner W 1980 *Z. Naturforsch. B: Chem. Sci.* **35** 1063
[27] Boolchand P, Robinson B L and Jha S 1970 *Phys. Rev. B* **2** 3463
[28] Poborchii V V, Kolobov A V and Tanaka K 1998 *Appl. Phys. Lett.* **72** 1167
[29] Poborchii V V, Kolobov A V and Tanaka K 1999 *Appl. Phys. Lett.* **74** 215
[30] Stephens R B 1984 *Phys. Rev. B* **30** 5195
[31] Hodge I M and Berens A R 2002 *Macromolecules* **15** 762
[32] Scherer G W 1990 *J. Non-Cryst. Solids* **123** 75
[33] Krüger J K, Alnot P, Baller J, Bactavatchalou R, Dorosz S, Henkel M, Kolle M, Krüger S P, Müller U, Philipp M, Possart W, Sanctuary R and Vergnat C 2007 About the Nature of the Structural Glass Transition: An Experimental Approach in *Ageing and the Glass Transition* (Berlin / New York: Springer)
[34] Boolchand P and Bresser W J 2000 *Phil. Mag. B* **80** 1757
[35] Micoulaut M 1998 *European Physical Journal B* **1** 277
[36] Kerner R and Micoulaut M 1997 *J. Non-Cryst. Solids* **210** 298
[37] Boolchand P, Micoulaut M and Chen P 2008 Nature of Glasses in *Phase Change Materials: Science and Applications* (Heidelberg: Springer)

[38]   Boolchand P, Georgiev D G and Micoulaut M 2002 *J. Optoelectron. Adv. Mater.* **4** 823
[39]   Saiter J M, Arnoult M and Grenet J 2005 *Physica B* **355** 370
[40]   Nagata K, Ishibashi K and Miyamoto Y 1981 *Jpn. J. Appl. Phys.* **20** 463
[41]   Feng X W, Bresser W J and Boolchand P 1997 *Phys. Rev. Lett.* **78** 4422
[42]   Mauro J C and Loucks R J 2008 *Phys. Rev. E* **78** 021502
[43]   Georgiev D G, Boolchand P, Eckert H, Micoulaut M and Jackson K 2003 *Europhys. Lett.* **62** 49
[44]   Maxwell R and Eckert H 1994 *J. Am. Chem. Soc.* **116** 682
[45]   Menard K P 2008 *Dynamic mechanical analysis : a practical introduction* (Boca Raton, FL: CRC Press)
[46]   Johari G 1987 Secondary relaxations and the properties of glasses and liquids in *Molecular Dynamics and Relaxation Phenomena in Glasses* (Berlin / Heidelberg: Springer)
[47]   Johari G P and Goldstein M 1970 *J. Chem. Phys.* **53** 2372
[48]   Nemilov S V 2001 *Glass Phys. Chem* **27** 214
[49]   Georgiev D G, Boolchand P and Jackson K A 2003 *Phil. Mag.* **83** 2941
[50]   Wachtman J, Chen P and Boolchand P 2009 *Bull. Am. Phys. Soc.* **54** 1001
[51]   Effey B and Cappelletti R L 1999 *Phys. Rev. B* **59** 4119
[52]   Chen P, Holbrook C, Boolchand P, Georgiev D G, Jackson K A and Micoulaut M 2008 *Phys. Rev. B* **78** 224208
[53]   Lucas P, King E A, Gueguen Y, Sangleboeuf J-C, Keryvin V, Erdmann R G, Delaizir G, Boussard-Pledel C, Bureau B, Zhang X-H and Rouxel T 2009 *J. Am. Ceram. Soc.* **92** 1986
[54]   Novita D I and Boolchand P 2007 *Phys. Rev. B* **76** 184205
[55]   Wilson M A, Carter M A, Hall C, Hoff W D, Ince C, Savage S D, Mckay B and Betts I M 2009 *Proc. R. Soc. A* **465** 2407
[56]   Shpotyuk O, Hyla M, Boyko V and Golovchak R 2008 *Physica B* **403** 3830
[57]   Boolchand P, Georgiev D G, Qu T, Wang F, Cai L C and Chakravarty S 2002 *Comptes Rendus Chimie* **5** 713
[58]   Boolchand P, Chen P and Vempati U 2009 *J. Non-Cryst. Solids* doi:10.1016/j.jnoncrysol.2008.11.046
[59]   Micoulaut M and Phillips J C 2003 *Phys. Rev. B* **67** 104204
[60]   Wang Y, Boolchand P and Micoulaut M 2000 *Europhys. Lett.* **52** 633
[61]   Mauro J C and Varshneya A K 2007 *J. Non-Cryst. Solids* **353** 1226
[62]   Parshin D A, Liu X, Brand O and Löhneysen H V 1993 *Z. Phys. B: Condens. Matter* **93** 57
[63]   Boolchand P, Feng X and Bresser W J 2001 *J. Non-Cryst. Solids* **293** 348
[64]   Selvanathan D, Bresser W J and Boolchand P 2000 *Phys. Rev. B* **61** 15061
[65]   Selvanathan D, Bresser W J, Boolchand P and Goodman B 1999 *Solid State Commun.* **111** 619
[66]   Massobrio C, Celino M and Pasquarello A 2003 *J. Phys. Condens. Matter* **15** S1537
[67]   Salmon P S, Martin R A, Mason P E and Cuello G J 2005 *Nature* **435** 75
[68]   Boolchand P, Chen P, Novita D I and Goodman B 2009 New perspectives on intermediate phases in *Rigidity transitions and Boolchand Intermediate Phases in nanomaterials* (Bucharest, Romania: INOE)
[69]   Massobrio C, Celino M, Salmon P S, Martin R A, Micoulaut M and Pasquarello A 2009 *Phys. Rev. B* **79** 174201

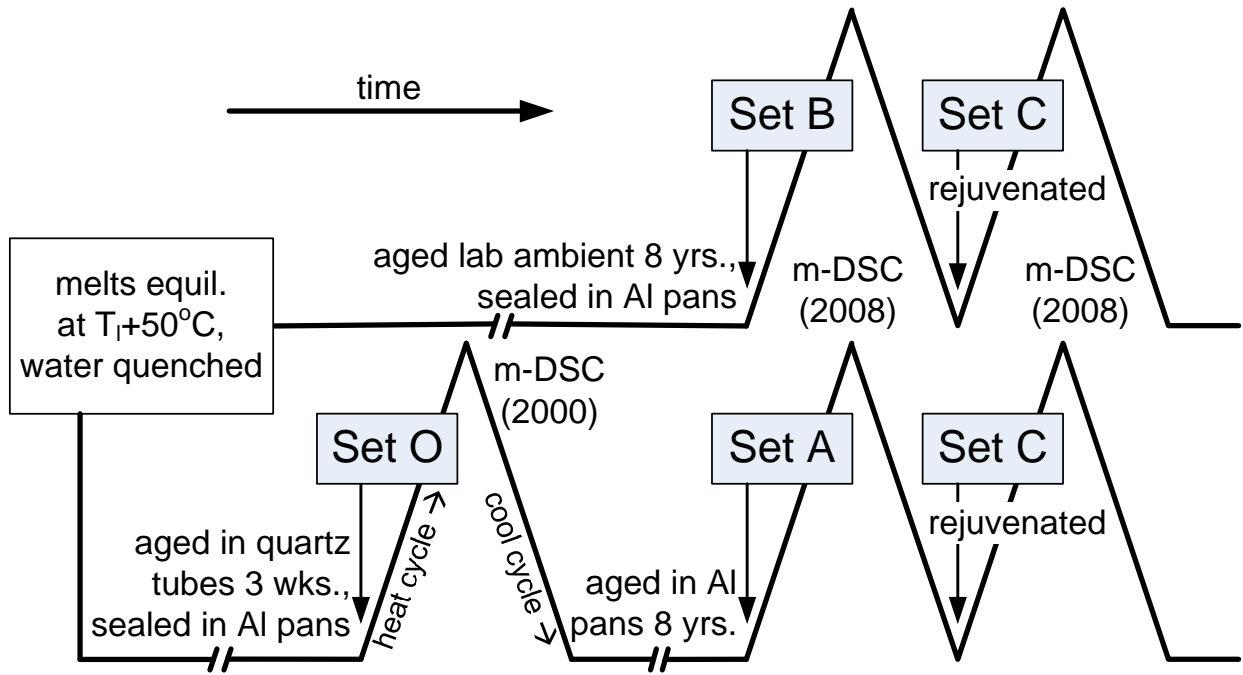

Ping Chen et al. - Figure 1

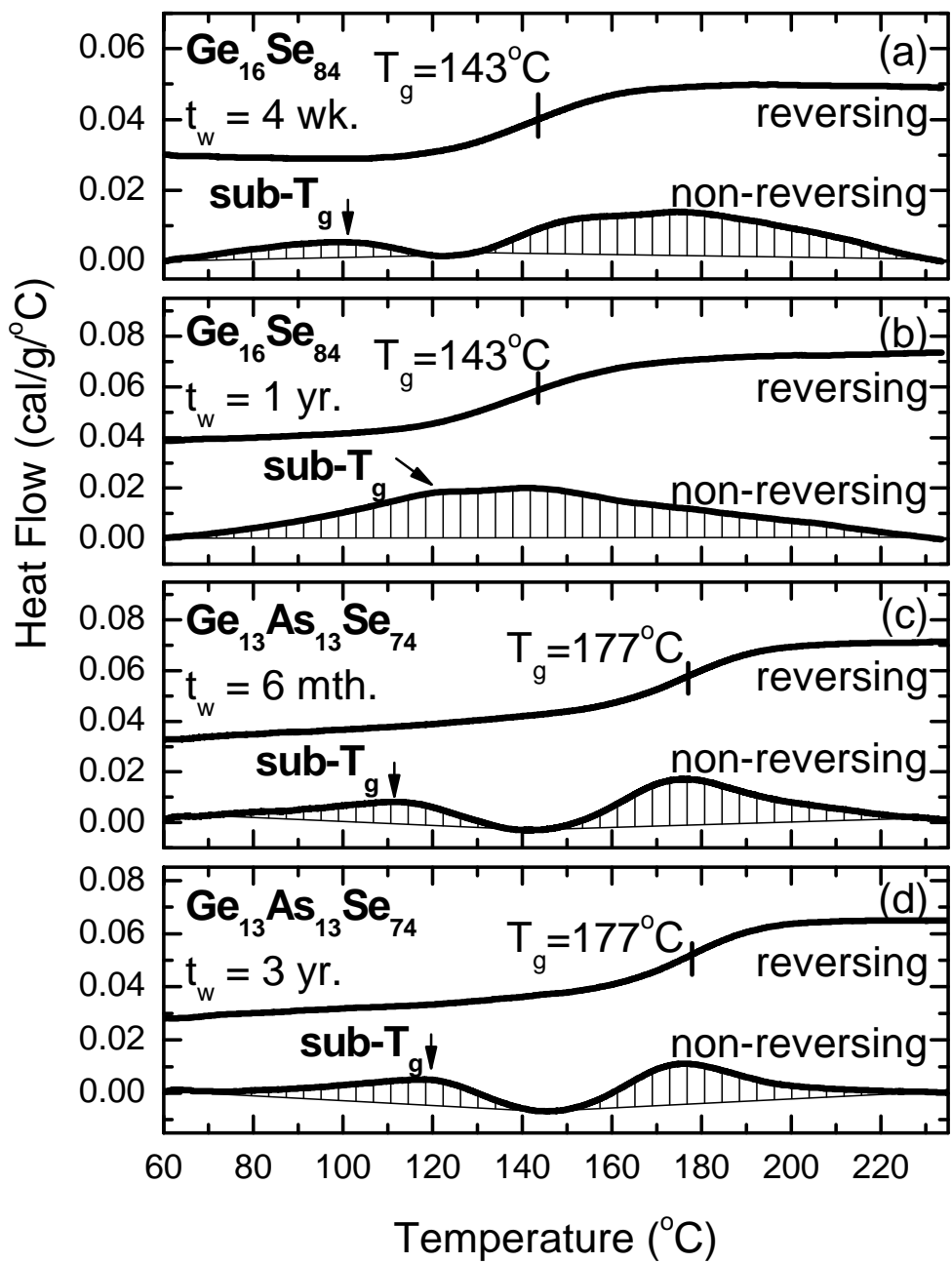

Ping Chen et al. - Figure 2

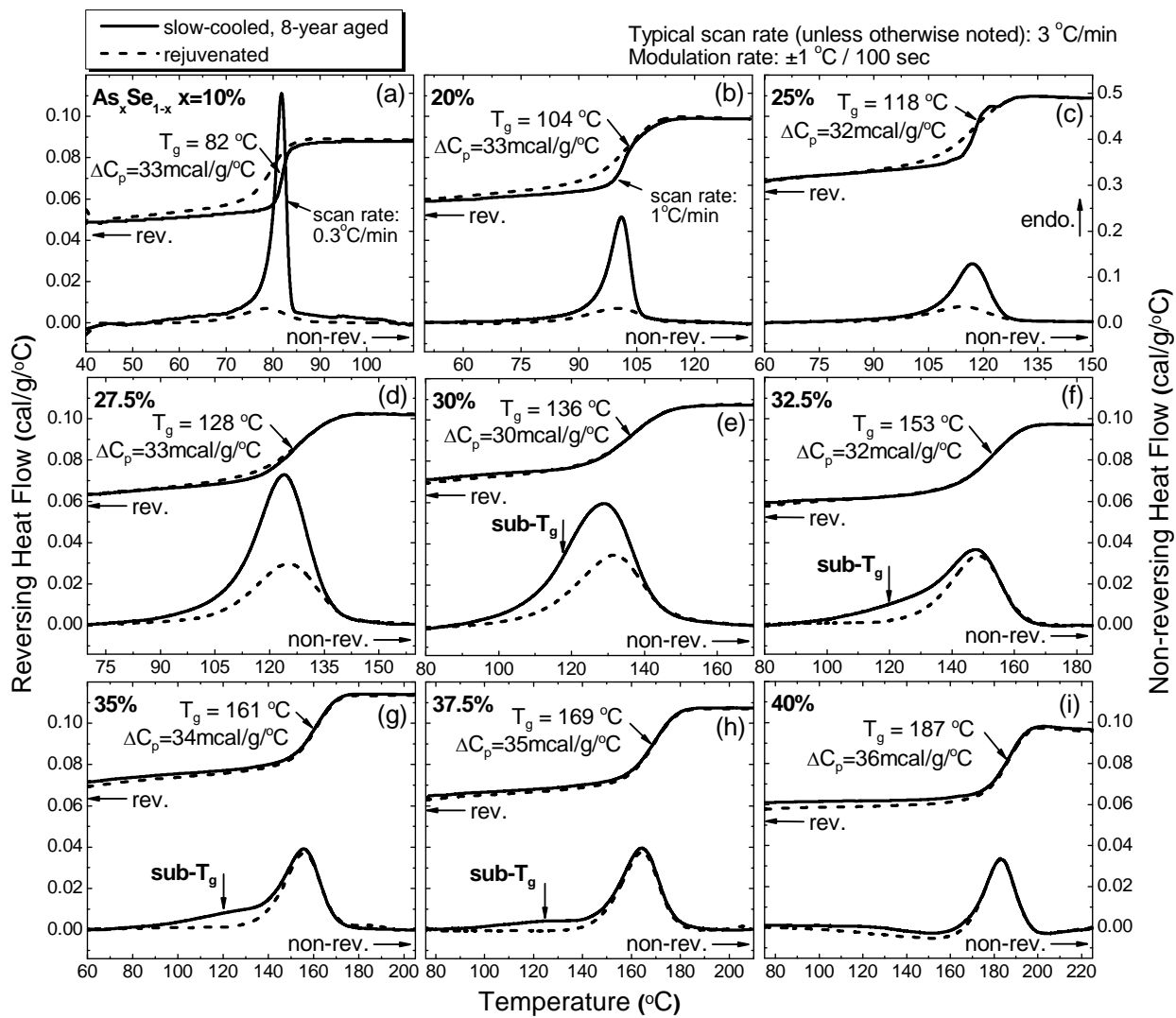

Ping Chen et al. - Figure 3

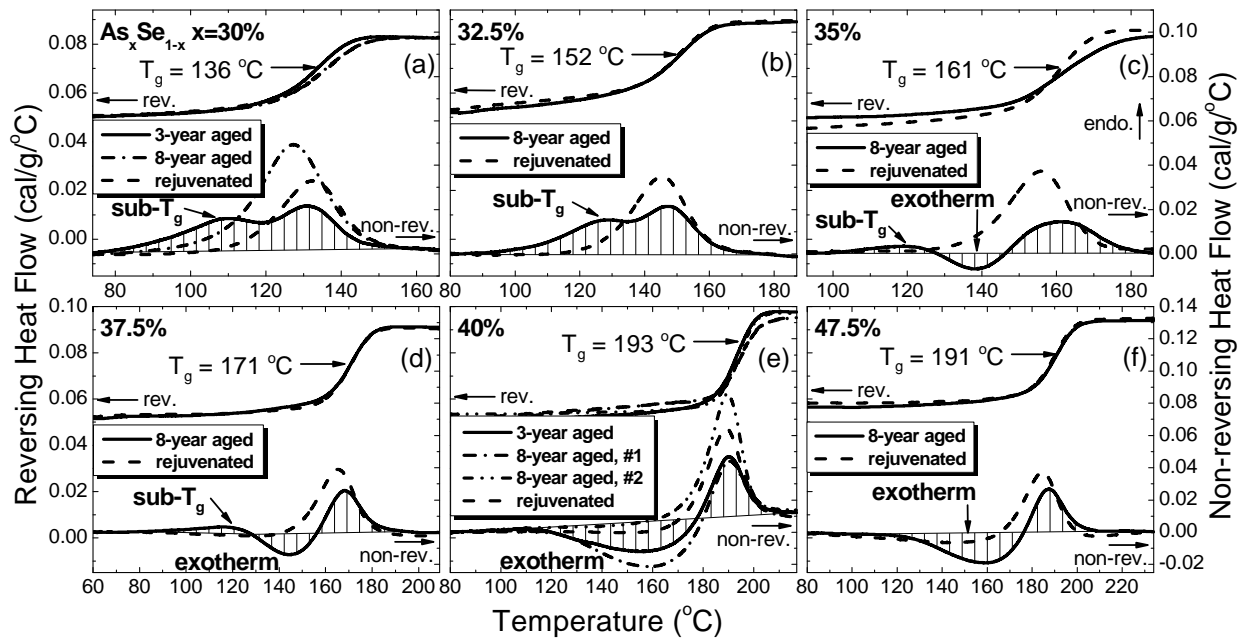

Ping Chen et al. - Figure 4

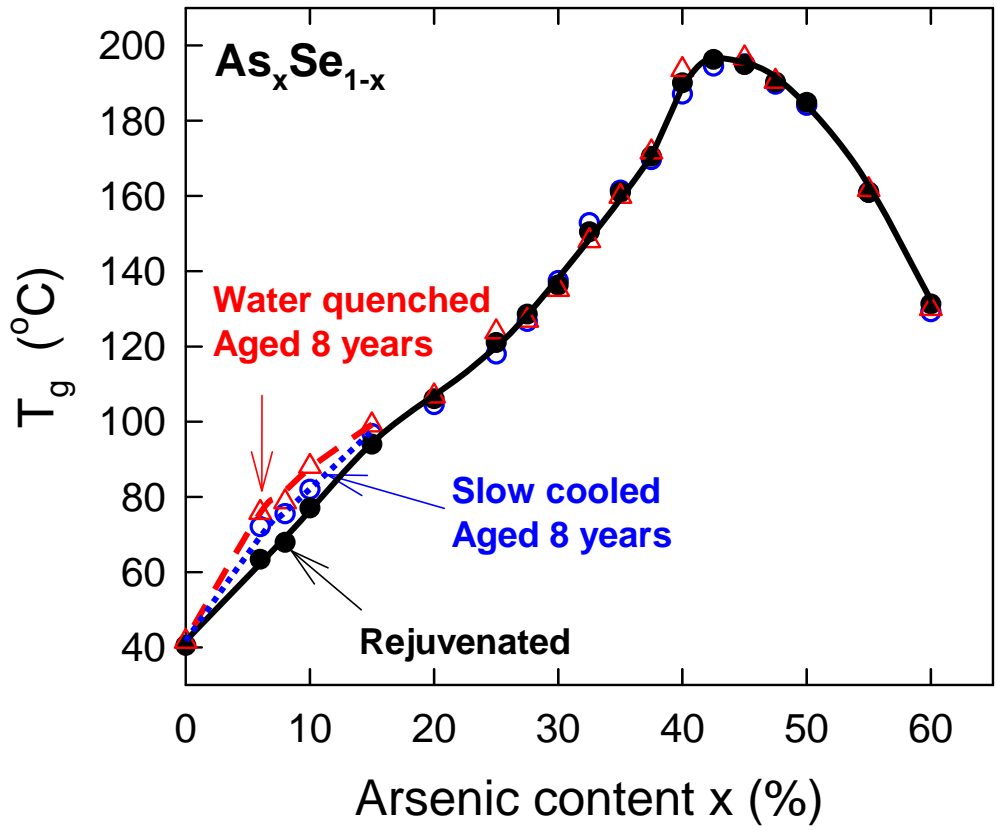

Ping Chen et al. - Figure 5

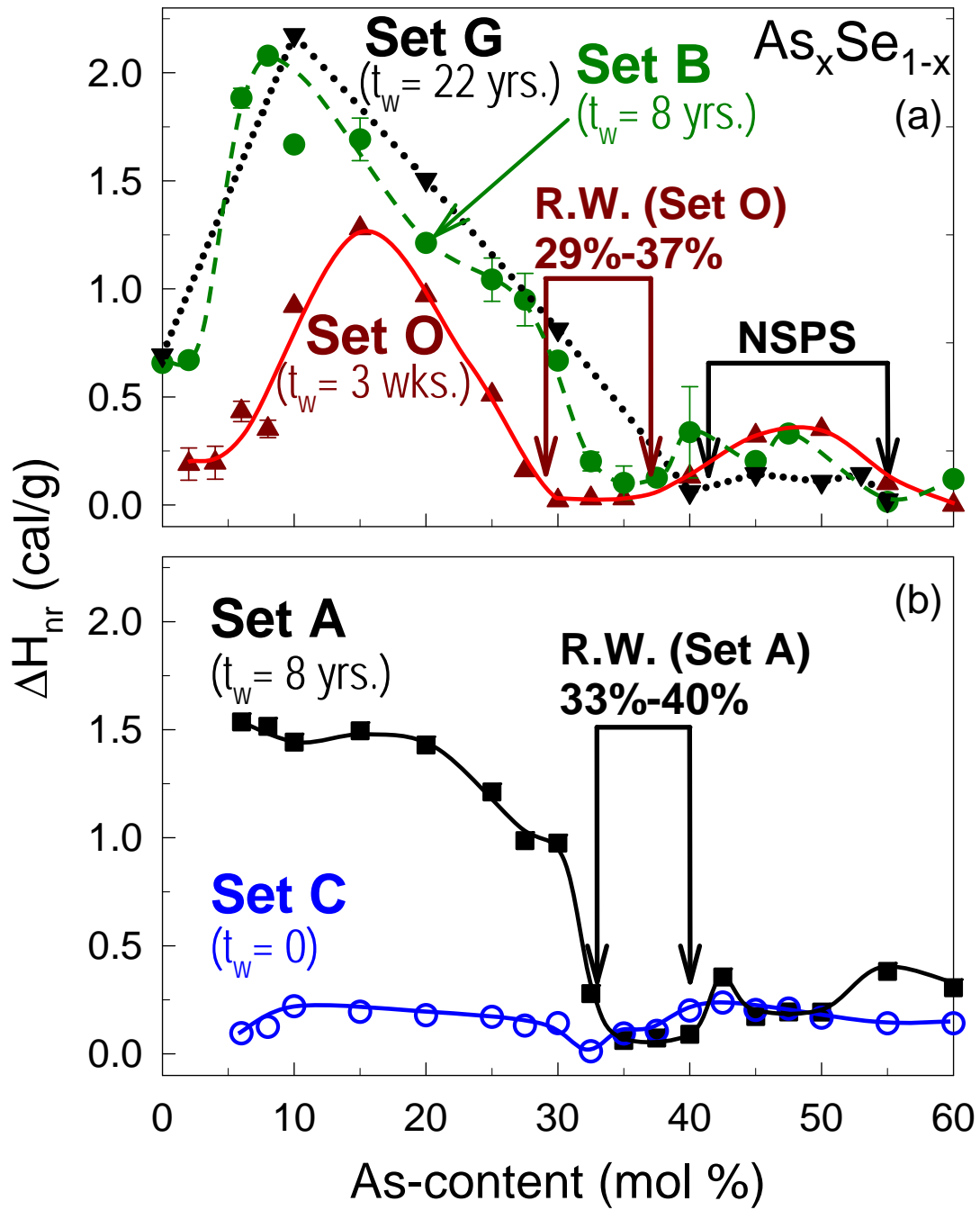

Ping Chen et al. - Figure 6

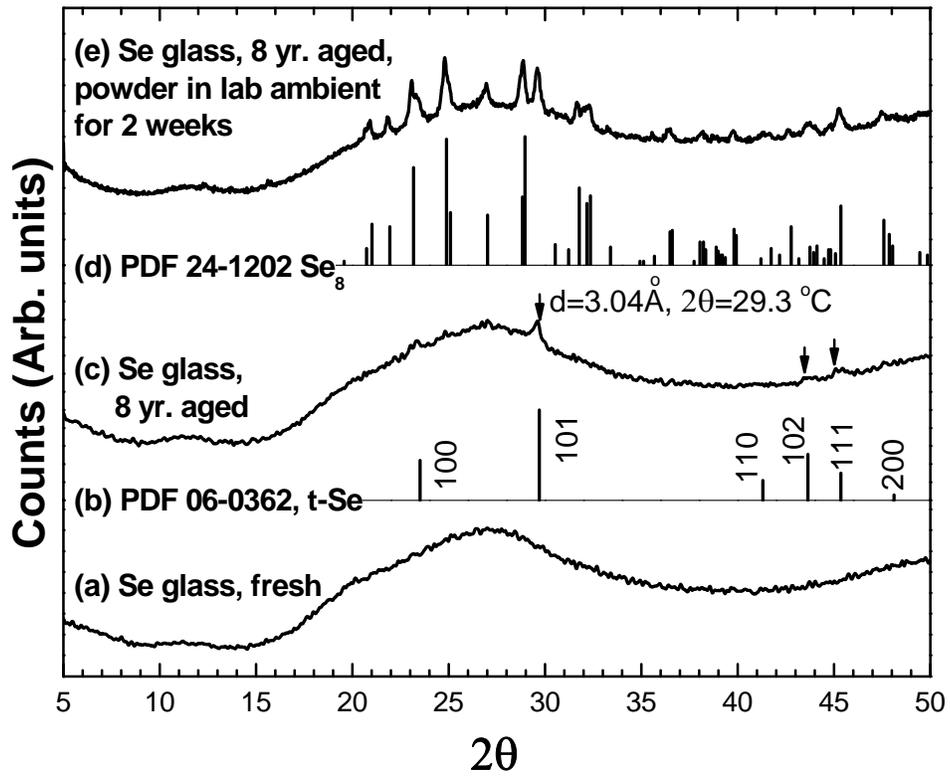

Ping Chen et al. - Figure 7

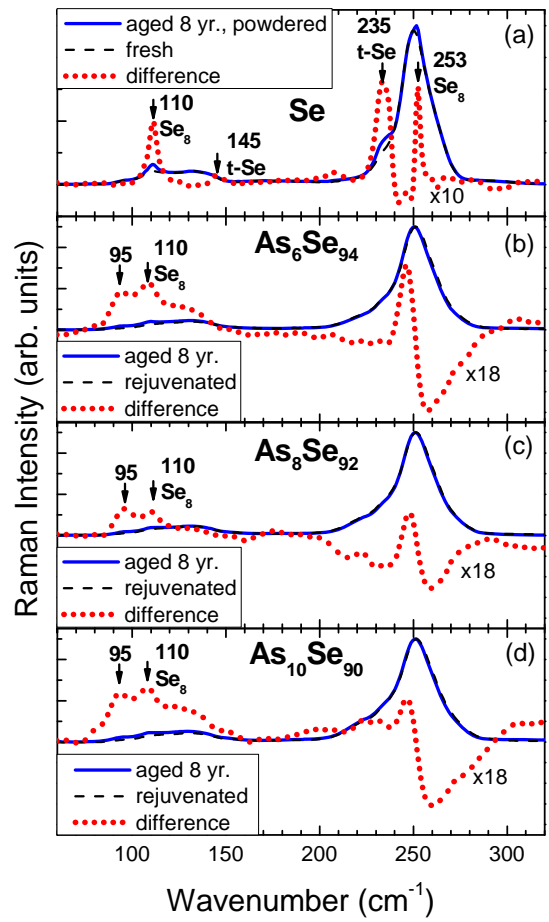

Ping Chen et al. - Figure 8

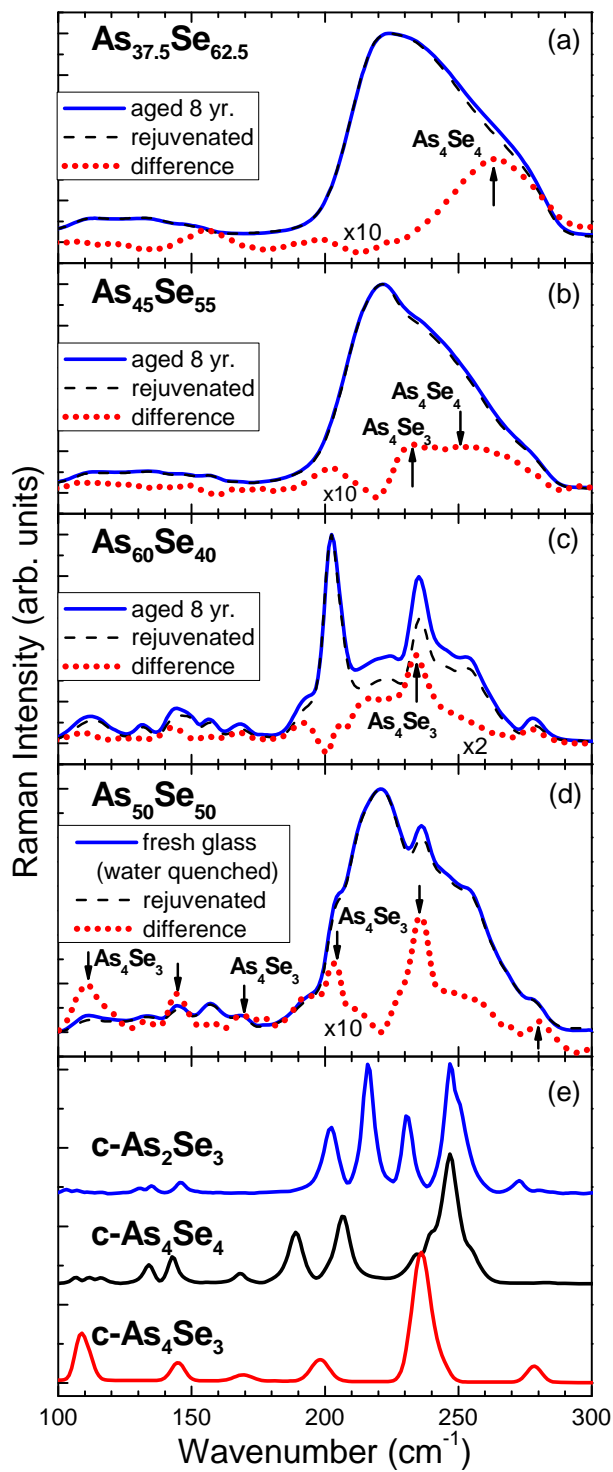

Ping Chen et al. - Figure 9

# Appendix I

**Enthalpy of Relaxation at $T_g$ from Modulated- DSC (MDSC) experiments.**

In this Appendix we provide a brief overview of mDSC as a technique to obtain the enthalpy of relaxation ($\Delta H_{nr}$) at $T_g$ in a glass. The term has been widely used to probe aging effects in the chalcogenides in the present work. The narrative below serves to introduce the method starting from DSC, the more familiar of the two thermal methods[19,20,30,31,32].

Differential Scanning Calorimetry (DSC) is widely used analytical method to investigate different types of thermal transitions/events including glass transition, melting, decomposition and crystallization. In standard DSC, a sample is heated at a linear temperature rate (ramp):

$$\bar{T}(t) = T_i + qt \qquad (A.1)$$

where $T_i$ is the initial temperature. The resulting heat flow has contributions from a heat capacity component and a kinetic component as below:

$$\dot{H}_{DSC}(t) = qC_p^{App} = qC_p + f(T,t) \qquad (A.2)$$

where $\dot{H}_{DSC}$ is the total heat flow rate (mW), $q$ is the linear heating rate (°C/min), $C_p^{App}$ is the apparent total heat capacity (J/°C), $C_p$ is the sample heat capacity (J/°C), and $f(T,t)$ is the kinetic part of the heat flow that is a function of both temperature(T) and time(t).

In standard DSC, one cannot separate heat capacity term from the kinetic term in a single scan. But it is possible to improve sensitivity and detect low energy transitions, by increasing heating rate ($q$) or sample weight. However, this will invariably lead to a loss of resolution, i.e., ability to resolve thermal events close in temperature, thus inhibiting optimizing both sensitivity and resolution simultaneously.

Temperature-Modulated DSC (MDSC) was developed in last decade [19,20] to solve this problem. In MDSC, a small sinusoidal temperature change $T_{sm}$ is added to the traditional DSC linear temperature ramp $\bar{T}(t)$:

$$\tilde{T}(t) = \bar{T}(t) + T_{sm} = T_i + qt + A\sin\omega t \tag{A.3}$$

where $T_i$ is the initial start temperature, $q$ is the underlying linear scan rate, $A$ is the amplitude of sinusoidal modulation and $\omega$ is the angular frequency of modulation. An example of MDSC temperature ramp is shown in Figure A1(a). The linear scan rate $q = 1$ K/min and the modulation amplitude $A = \pm 1$ K over a period of 100 sec ($\omega = 0.0628$ Hz). We used a TA Instruments, model MDSC 2920 with a refrigerated cooling system (RCS) for investigating the glass transition in Se glass. The programmed modulated temperature ramp rate becomes:

$$\tilde{q}(t) = \frac{d\tilde{T}(t)}{dt} = q + A\omega\cos\omega t \tag{A.4}$$

As in DSC, the modulated heat flow rate generated in MDSC will also have a heat capacity component and a kinetic component:

$$\dot{H}_{MDSC}(t) = C_p \tilde{q}(t) + f(T,t)$$
$$= qC_p + A\omega C_p \cos(\omega t - \varphi) + f(T,t) \tag{A.5}$$

where $\varphi$ is the phase lag angle. In MDSC the heat flow signal consists of three parts: heat capacity component in response to linear T ramp, i.e., the *Reversing Heat Flow* $\dot{H}_{Rev.}$; $C_p$ part in response to sinusoidal T-modulation which is the amplitude of the *Modulated Heat Flow Amplitude* $\dot{H}_0$; and finally the kinetic component which represents the Non-reversing heat flow $\dot{H}_{Nonrev.}$. We can rewrite equation (A.5) as:

$$\dot{H}_{MDSC}(t) = \left[\dot{H}_{Rev.} + \dot{H}_{Nonrev.}\right] + \dot{H}_0 \cos(\omega t - \varphi)$$
$$\dot{H}_{Rev.} \equiv qC_p, \quad \dot{H}_{Nonrev.} \equiv f(T,t), \quad \dot{H}_0 \equiv A\omega C_p \tag{A.6}$$

and define the sum of $\dot{H}_{Rev.}$ and $\dot{H}_{Nonrev.}$ as *Total Heat Flow* $\dot{H}_{Total}$, a quantity which is exactly the same as observed in a standard DSC experiment, $\dot{H}_{DSC}$ (see equation A.2). So now we get:

$$\dot{H}_{MDSC}(t) = \dot{H}_{Total} + \dot{H}_0 \cos(\omega t - \varphi), \quad \dot{H}_{Total.} \equiv \dot{H}_{Rev.} + \dot{H}_{Nonrev.} \tag{A.7}$$

From the measured $\dot{H}_{MDSC}$ signal, we can accurately extract the average as $\langle \dot{H}_{MDSC} \rangle$ and amplitude $\dot{H}_0$ using Fourier transformation as shown in Fig A1 (b). The heat capacity can be obtained as follows:

$$C_p = \left(\frac{K_{Cp}}{A\omega}\right)\dot{H}_0 \tag{A.8}$$

where $K_{Cp}$ is a calibration constant of the MDSC cell usually performed using a sapphire standard. In summary, we have a set of 3 heat-flow signals as below:

$$\begin{cases} \dot{H}_{Total} = \langle \dot{H}_{MDSC} \rangle \\ \dot{H}_{Rev.} = qC_p = \left(\dfrac{qK_{Cp}}{A\omega}\right)\dot{H}_0 \\ \dot{H}_{Nonrev.} = \dot{H}_{Total} - \dot{H}_{Rev.} \end{cases} \tag{A.9}$$

In Figure A.1(c) we have display these signals. The *Total Heat Flow* signal is the average value of the measured modulated heat flow ($\dot{H}_{MDSC}$) signal, the *Reversing Heat Flow* signal is the *Heat Flow Amplitude* $\dot{H}_0$ signal, and finally the *Non-reversing Heat flow* signal is obtained by subtracting $\dot{H}_{Total}$ from $\dot{H}_{Rev.}$. In these MDSC experiments the modulation amplitude, modulation frequency and linear heating rate are kept constant. The inflection point of the rounded step observed in the reversing heat flow signal $\dot{H}_{Rev.}$ is used to define the glass transition temperature ($T_g$), and for Se glass we obtain a $T_g$ close to 40°C (Figure A.1(c)). The non-reversing heat flow signal reveals a peak as a precursor to the glass transition, and the integrated area under the peak provides a measure of the enthalpy of relaxation or the non-reversing enthalpy ($\Delta H_{nr}$) at $T_g$.

**Modulation Frequency corrected Non-reversing enthalpy at $T_g$ from MDSC experiments.**

The *non-reversing heat* flow signal equals the difference between the total $\dot{H}_{Total}$ and reversing heat flow signal $\dot{H}_{Rev.}$ (equation A9). The *reversing heat* flow signal generally upshifts in T as the modulation frequency ($\omega$) increases, while the total heat flow signal is independent of $\omega$ [19]. For these reasons, the *non-reversing enthalpy* ($\Delta H_{nr}$) measured in a heating cycle across $T_g$ is modulation frequency dependent.

To obtain the frequency corrected *non-reversing enthalpy* ($\Delta H_{nr}$) in an MDSC experiment, it is usual to program a cooling cycle following the heating cycle across $T_g$. The T-upshift of the *reversing heat* flow signal in the heating cycle leads to an overestimate of the *non-reversing enthalpy* term. The overestimate is obtained directly from the cooling cycle wherein the *reversing heat flow* signal now downshifts in T by exactly the same amount as it had upshifted in the heating cycle, and leads to a <u>finite</u> non-reversing enthalpy. Figure A2 shows an MDSC scan of a $As_{27.5}Se_{72.5}$ bulk glass. The non-reversing heat flow signal in the cooling cycle shows an exotherm, and this term provides exactly the overestimate in $\Delta H_{nr}$ one was looking for in the heating cycle. Thus, the procedure consists of subtracting the *non-reversing enthalpy* (exotherm) of 0.30 cal/gm measured in the cooling cycle from the *non-reversing enthalpy* (endotherm) measured in the heating cycle of 1.11 cal/gm, and yields the *frequency corrected non-reversing enthalpy ($\Delta H_{nr}$)* of 0.81(10) cal/gm for the sample in question. We have routinely followed this procedure to obtain the frequency corrected non-reversing enthalpy in our MDSC experiments.

A second procedure to estimate the frequency corrected non-reversing heat flow is to carry forward a third cycle, an additional heating scan (cycle #3) immediately following the two cycles (#1 and #2) mentioned above. Cycle #3 permits recording the enthalpy of relaxation of the *rejuvenated* sample, i.e., a sample in which the aging time clock is reset to zero. The difference in the *non-reversing enthalpy* between cycles #1 and #3 provides a direct measure of the enthalpy of relaxation upon aging- the quantity of interest here. While not as accurate as the first procedure described above, the rejuvenation process does provide an independent means to ascertain general trends in the variation of the non-reversing enthalpy on account of aging.

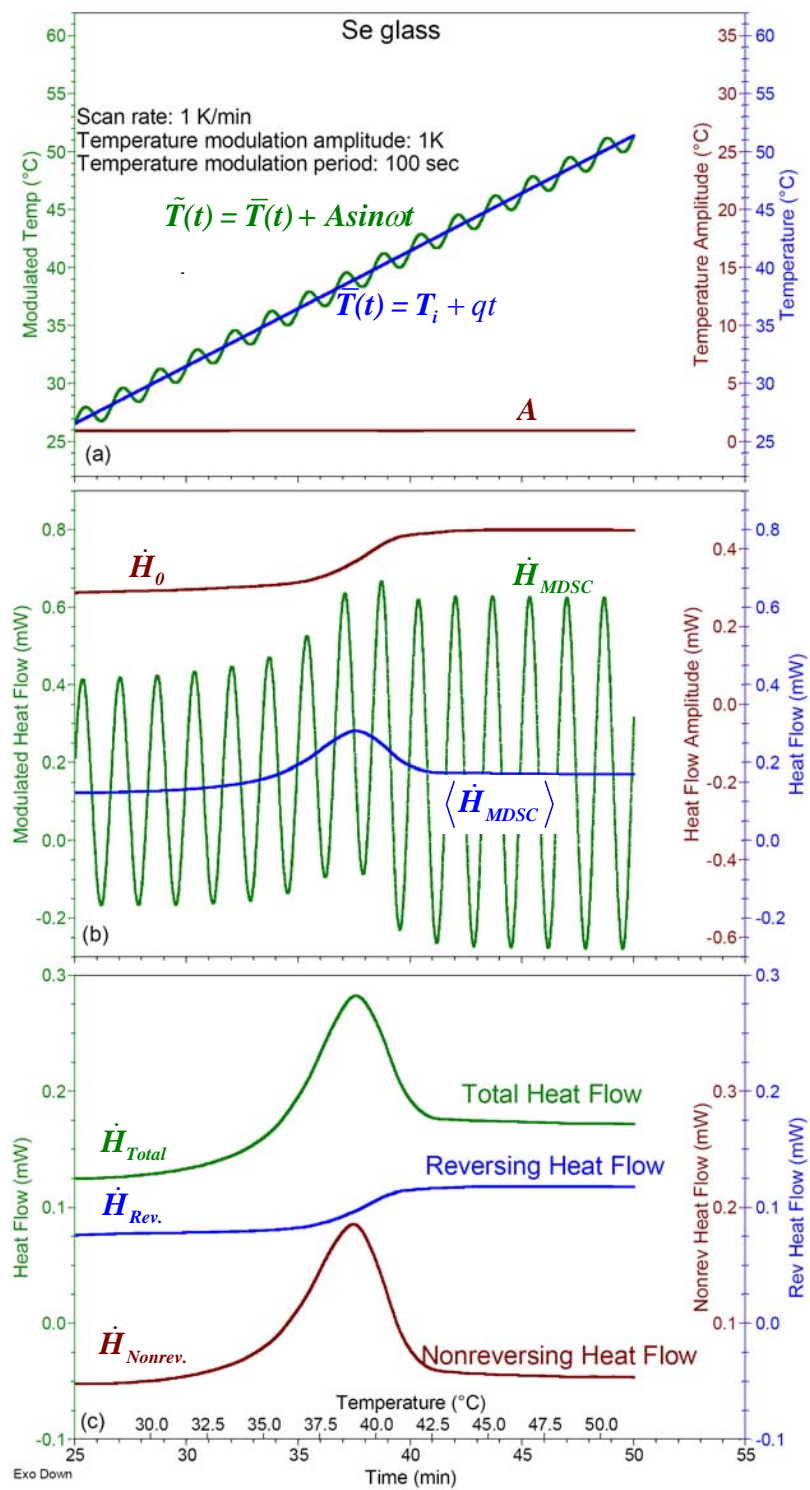

**Figure A1**. Panel (a) Programmed T-profile used for mDSC scans. (b) Modulated heat flow in green curve, the average of the modulated heat flow representing the total heat flow as blue curve, and the amplitude of the modulations representing the reversing heat flow as a brown curve (c). The total heat flow (green), reversing heat flow (blue) and the non-reversing heat flow (brown) curves for Se glass.

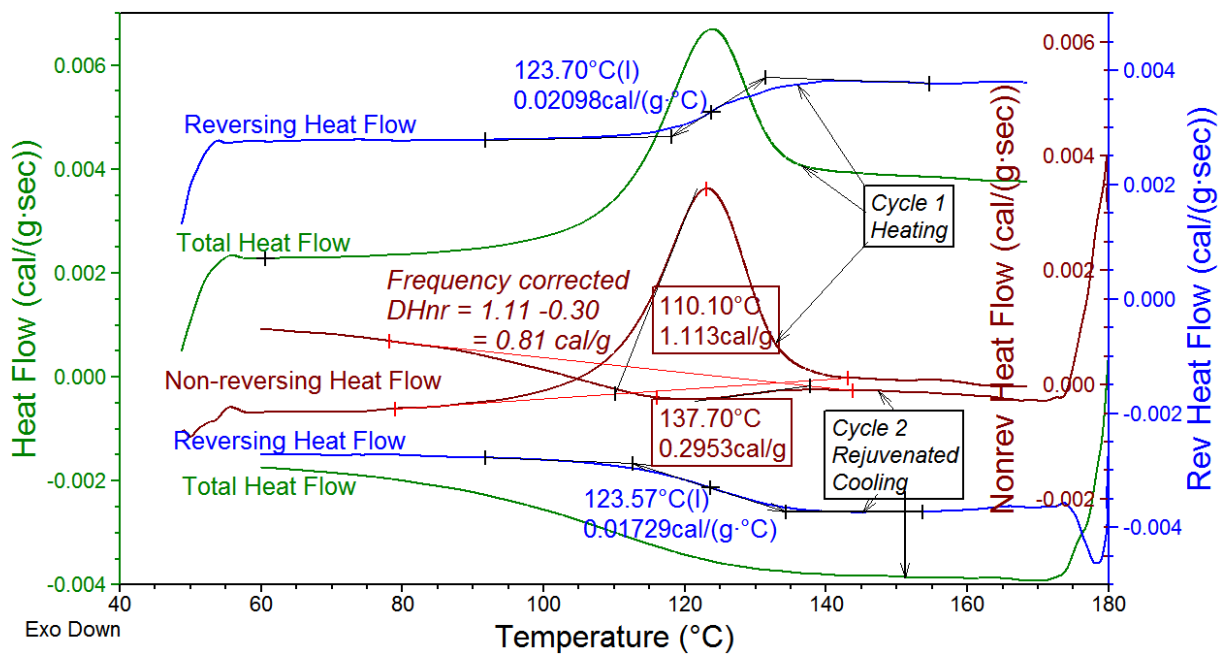

Figure A2. Modulated DSC scan of $As_{27.5}Se_{72.5}$ glass taken at 3°C/min scan rate and 1/100sec modulation rate showing the total, reversing and non-reversing heat flow signals in the heating cycle (cycle 1,top three) and in the cooling cycle (cycle 2,bottom three). The Frequency corrected non-reversing enthalpy of the glass is found to be 0.81 cal/gm. See text for details.